\documentclass[aps,amsmath,amssymb,nofootinbib,
reprint,longbibliography,superscriptaddress,
preprintnumbers,prd]{revtex4-2}
\pdfoutput=1
\usepackage{graphicx}
\usepackage{subcaption}
\usepackage{lipsum}
\usepackage{booktabs}
\usepackage{tabularx}
\usepackage{tikz}
\usepackage{pgfplots}
\usepackage{comment}
\usetikzlibrary{positioning,shapes,trees,arrows,graphs,decorations,calc}
\usepackage{url}

\usepackage{amssymb}
\usepackage{amsmath}
\usepackage{hyperref}
\usepackage{xcolor}

\usepackage{placeins} 
\usepackage[normalem]{ulem}
\usepackage{xspace}
\usepackage{lipsum}

\usepackage{cellspace} %
\newcommand{\eq}[1]{eq.~\eqref{eq:#1}}

\newcommand{\eqs}[2]{eqs.~\eqref{eq:#1} and \eqref{eq:#2}}
\renewcommand{\sec}[1]{sec.~\ref{sec:#1}}

\newcommand{\app}[1]{app.~\ref{app:#1}}

\newcommand{\refcite}[1]{ref.~\cite{#1}}
\newcommand{\Refcite}[1]{Ref.~\cite{#1}}
\DeclareRobustCommand{\refscite}[1]{refs.~\cite{#1}}

\newcommand{\ord}[1]{\mathcal{O}(#1)}

\newcommand{\df}{\mathrm{d}}

\newcommand{\as}{\alpha_{s}}

\newcommand{\Ecm}{E_{\mathrm{cm}}}

\newcommand{\Tau}{\mathcal{T}}

\newcommand{\TaupT}{\mathcal{T}_1^{p_T}}

\newcommand{\GeV}{\,\mathrm{GeV}}
\newcommand{\TeV}{\,\mathrm{TeV}}

\newcommand{\nn}{\nonumber}

\newcommand{\cut}{\mathrm{cut}}

\newcommand{\FO}{\mathrm{FO}}

\newcommand{\NtLL}{\mathrm{N}^3\mathrm{LL}}

\newcommand{\sing}{\mathrm{sing}}

\newcommand{\res}{\mathrm{res}}

\newcommand{\cusp}{\mathrm{cusp}}

\newcommand{\frun}{f_\mathrm{run}}
\newcommand{\grun}{g_\mathrm{run}}
\newcommand{\vary}{\mathrm{vary}}
\newcommand{\run}{\mathrm{run}}
\newcommand{\muZeromin}{\mu_0^\mathrm{min}}

\newcommand{\mufmin}{\mu_f^\mathrm{min}}

\newcommand{\lqcd}{\Lambda_\mathrm{QCD}}

\newcommand{\geneva}{\textsc{Geneva}\xspace}
\newcommand{\radish}{\textsc{RadISH}\xspace}

\newcommand{\pythiaEight}{\textsc{Pythia8}\xspace}

\newcommand{\scetlib}{\textsc{SCETlib}\xspace}

\makeatletter
\g@addto@macro\bfseries{\boldmath}
\makeatother

\DeclareMathAlphabet\mathbfcal{OMS}{cmsy}{b}{n}

\begin{document}


\title{Resumming transverse observables for NNLO+PS matching in GENEVA}

\author{Alessandro Gavardi}
\affiliation{Deutsches Elektronen-Synchrotron DESY, Notkestr. 85, 22607 Hamburg, Germany}

\author{Rebecca von Kuk}
\affiliation{Deutsches Elektronen-Synchrotron DESY, Notkestr. 85, 22607 Hamburg, Germany}

\author{Matthew~A.~Lim}
\affiliation{Department of Physics and Astronomy, University of Sussex, Sussex House, Brighton, BN1 9RH, UK}
\affiliation{Universit\`a degli Studi di Milano-Bicocca \& INFN Sezione di Milano-Bicocca, Piazza della Scienza 3, Milano 20126, Italy}

\preprint{DESY-25-068}

\date{\today}

\begin{abstract}
We study the use of higher-order resummation for transverse observables to achieve NNLO+PS matching within the \geneva framework. In particular, we embed $q_T$ resummation for colour-singlet production at N$^3$LL obtained via soft-collinear effective theory and implemented in the library \scetlib within \geneva. We also study for the first time the use of the generalised $N$-jettiness variable in parton shower matching, and achieve the resummation of the one-jettiness defined with transverse measures up to NLL$'$ accuracy. As a case study, we use these resummed calculations to construct a \geneva NNLO+PS generator for Higgs boson production in heavy-quark annihilation (with beauty or charm-quarks in the initial state). The use of transverse measures facilitates the matching to showers ordered in transverse momentum, and opens the door to possible future extensions of this approach to the production of colour singlets in association with final-state jets.
\end{abstract}

\maketitle

\section{Introduction}

The success of the experimental programme at the LHC demands theoretical tools at the highest possible accuracy. While fully-differential calculations at fixed-order in perturbation theory have reached N$^3$LO for simple processes~\cite{Billis:2021ecs,Chen:2021isd,Chen:2022cgv,Chen:2025kez}, state-of-the-art for predictions matched to parton shower generators is currently next-to-next-to-leading order (NNLO+PS). Two main approaches exist to achieve this accuracy~\cite{Alioli:2012fc,Monni:2019whf}, while a third, rather different approach, is currently under active development~\cite{Campbell:2021svd,El-Menoufi:2024sys}.

An interesting question relevant to the construction of NNLO+PS generators concerns the choice of resolution variables, which partition the phase space into jet bins of differing multiplicity and which must be resummed to high logarithmic accuracies. While many choices are in principle possible, the predictions for exclusive observables will depend on said choice, introducing a source of systematic uncertainty. A natural way to gauge the size of this uncertainty would be to compare the predictions of generators constructed using the same method but different variables. In addition, the availability of higher-accuracy parton shower algorithms~\cite{Dasgupta:2020fwr,vanBeekveld:2024wws,Herren:2022jej,Hoche:2024dee} means that one may wish to explore novel variable choices to ensure the preservation of the shower accuracy in matching.

In this work, we implement a new combination of resolution variables in the \geneva approach. The original formulation of \geneva~\cite{Alioli:2015toa} used the zero- and one-jettiness~\cite{Stewart:2010tn} to separate the 0/1- and 1/2- jet bins respectively, and indeed many electroweak processes have since been examined in this context~\cite{Alioli:2019qzz, Alioli:2020fzf,Alioli:2020qrd,Alioli:2021egp, Cridge:2021hfr,Alioli:2022dkj,Alioli:2023har}. The jettiness variable has the advantage of considerable simplicity, admitting a simple factorisation structure in SCET-I which facilitates resummation~\cite{Stewart:2009yx,Alioli:2021ggd}. However, the fact that the usual definition of the jettiness involves an invariant mass-like measure complicates the matching to commonly employed parton shower algorithms, which tend to be ordered in transverse momentum. An alternative formulation using the colour-singlet transverse momentum $q_T$ to separate the 0/1-jet bins was studied in \refcite{Alioli:2021qbf}, with $q_T$ resummation at N$^3$LL provided by the \radish formalism~\cite{Monni:2016ktx,Bizon:2017rah}. The 1/2-jet separation variable remained, however, the one-jettiness. In \refcite{Gavardi:2023aco}, a generator for $W^+W^-$ production was constructed using the hardest and second-hardest jet transverse momenta as variables and SCET-based resummation~\cite{Stewart:2013faa,Cal:2024yjz}. With these choices, the need for truncated showering techniques~\cite{Nason:2004rx} in \geneva was reduced. A summary of the currently available \geneva implementations is presented in tab.~\ref{tab:flavours}.

The new class of generators we initiate in this work uses transverse
observables to separate all jet bins, as in
\refcite{Gavardi:2023aco}. As in \refcite{Alioli:2021qbf}, we use
$q_T$ as the primary resolution variable, though exploiting analytic
SCET-II resummation provided by \scetlib~\cite{scetlib} rather than
the \radish approach previously taken. As a secondary resolution
variable, we choose a generalisation of the one-jettiness which uses a
transverse momentum-like measure. Though the definition of this
observable was introduced in the original $N$-jettiness paper over 15
years ago~\cite{Stewart:2010tn}, to our knowledge this is the first
time that its resummation has been accomplished beyond leading
logarithmic order.~\footnote{The NLL resummation should also be
achievable within the CAESAR formalism~\cite{Banfi:2004yd}, but to our
knowledge, this observable has never been studied in that
context.}~\footnote{We note that a similar generalisation of the
jettiness has more recently been proposed in
\refcite{Buonocore:2022mle}.} The observable is interesting not merely
because it facilitates matching for colour singlet processes, but
because of the promise it holds for achieving NNLO+PS matching for
colour-singlet production in association with additional jets. The
work we present here takes a very first step in this direction,
achieving NLL$'$ resummation before the shower matching.

As a case study, we consider the production of a Higgs boson through heavy-quark annihilation ($c\bar{c}, b\bar{b}\to H$). The requisite $q_T$ resummation for this process was achieved at N$^3$LL in \refcite{Cal:2023mib}, and generators for the $b\bar{b}H$ process were constructed in \refscite{Biello:2024vdh,Biello:2024pgo} using the \textsc{MiNNLOps} formalism and approximations for unknown two-loop matrix elements. As detailed in e.g. \refcite{Bishara:2016jga}, measurements of the Higgs boson $q_T$ spectrum could be used in conjunction with precision theory predictions for $b\bar{b}H$ to extract a value for the bottom Yukawa coupling $y_b$. Generalisations to other colour-singlet processes, though time-consuming, are in principle straightforward.

The rest of this paper is arranged as follows. In section~\ref{sec:geneva}, we provide a brief recap of the \geneva method for matching NNLO calculations to parton shower, while section~\ref{sec:scetqt} summarises $q_T$ resummation in SCET. In section~\ref{sec:tau1pt} we briefly review the definition of the generalised jettiness for unacquainted readers, before detailing the resummation of a specific choice of measures at NLL$'$. We present NNLO+PS matched results for $b\bar{b}H$ and $c\bar{c}H$ in section~\ref{sec:results}, before concluding in section~\ref{sec:conc}.

\section{The GENEVA method}
\label{sec:geneva}
The \geneva method \cite{Alioli:2012fc, Alioli:2015toa} relies on
defining infrared (IR) safe events at a specific perturbative order,
which are obtained by combining fixed-order and resummed calculations.
This is achieved by converting IR-divergent final states with $M$
partons into IR-finite final states with $N$ partonic jets, where $M
\geq N$, ensuring that the divergences cancel on an event-by-event
basis.  The conversion is performed using $N$-jet resolution variables
$r_N$ to define slicing parameters $r_N^\cut$, which divide the phase
space into regions with different numbers of resolved emissions:
$\Phi_0$ where there is no additional jet, $\Phi_1$ with one jet and
$\Phi_2$ with two or more jets in the final state.

To define the \geneva differential cross sections with zero, one and
two jets, we begin by introducing the following short-hand notation
\begin{align}
  \frac{\df \sigma^{\mathrm{NLO}_{0}}}{\df \Phi_0}(r_0^{\mathrm{cut}})
  =& B_0(\Phi_0)+ V_0(\Phi_0)
  \nn \\
  & + \int \frac{\df \Phi_1}{\df \Phi_0} \, B_1(\Phi_1) \,
  \theta(r_0<r_0^{\mathrm{cut}}),
  \\
  \frac{\df \sigma^{\mathrm{NLO}_{1}}}{\df \Phi_2}(r_1^{\mathrm{cut}})
  =& B_1(\Phi_1)+ V_1(\Phi_1)
  \nn \\
  & + \int \frac{\df \Phi_2}{\df \Phi_1} \, B_2(\Phi_2) \,
  \theta(r_1<r_1^{\mathrm{cut}}),
  \\
  \frac{\df \sigma^{\mathrm{LO}_{2}}}{\df \Phi_2} =& B_2(\Phi_2).
\end{align}
where $B_M$ and $V_M$ represent the tree-level and one-loop
contributions with $M$ partons in the final state, and we defined
\begin{align}
  \frac{\df \Phi_{N+1}}{\df \Phi_N} = \df \Phi_{N+1} \, \delta[\Phi_N
    - \tilde{\Phi}_N(\Phi_{N+1})]
\end{align}
to indicate the integration over the $\Phi_{N+1}$ phase space points
that are projected on $\Phi_N$ by the $\tilde{\Phi}_N$
mapping. Following the method first developed in
\refcite{Gavardi:2023aco}, the differential cross sections with
zero, one and two additional jets are then defined as

\begin{widetext}
  \begin{align}
    \label{eq:0-jet_cross_section}
    \frac{\df \sigma_0^\mathrm{MC}}{\df \Phi_0}(r_0^\mathrm{cut}) = &
    \frac{\df \sigma^{\NtLL_{r_0}}}{\df \Phi_0}(r_0^\mathrm{cut}) -
    \frac{\df \sigma^{\NtLL_{r_0}}}{\df \Phi_0}(r_0^\mathrm{cut})
    \Bigg \vert_{\mathrm{NLO}_0} + \frac{\df
      \sigma^{\mathrm{NLO}_0}}{\df \Phi_0}(r_0^\mathrm{cut})
    \\
    \label{eq:1-jet_cross_section}
    \frac{\df \sigma_1^\mathrm{MC}}{\df \Phi_1}(r_1^\mathrm{cut}) = &
    \Bigg \{\left[\frac{\df \sigma^{\NtLL_{r_0}}}{\df \Phi_0 \, \df
        r_0}-\frac{\df \sigma^{\NtLL_{r_0}}}{\df \Phi_0 \, \df
        r_0}\Bigg \vert_{\mathrm{NLO}_1}\right] \mathcal{P}_{0 \to
      1}(\Phi_1) \, U_1(\Phi_1, r_1^{\mathrm{cut}})
    \nn \\
    &+\frac{\df \sigma^{\mathrm{NLO}_1}}{\df \Phi_1}(r_1^\mathrm{cut})
    +\frac{\df \sigma^{\mathrm{NLL}'_{r_1}}}{\df
      \Phi_1}(r_1^\mathrm{cut}) -\frac{\df
      \sigma^{\mathrm{NLL}'_{r_1}}}{\df \Phi_1}(r_1^\mathrm{cut})\Bigg
    \vert_{\mathrm{NLO}_1} \Bigg \}\,\theta(r_0>r_0^\mathrm{cut})
    \nn \\
    & + \frac{\df \sigma_\mathrm{nonproj}^{\mathrm{LO}_1}}{\df \Phi_1}
    \, \theta(r_0<r_0^\mathrm{cut})
    \\
    \label{eq:2-jet_cross_section}
    \frac{\df \sigma_{\geq 2}^\mathrm{MC}}{\df \Phi_2} = & \Bigg \{
    \left[\frac{\df \sigma^{\NtLL_{r_0}}}{\df \Phi_0 \, \df r_0}
      -\frac{\df \sigma^{\NtLL_{r_0}}}{\df \Phi_0 \, \df r_0}\Bigg
      \vert_{\mathrm{NLO}_1}\right] \mathcal{P}_{0 \to 1}(\Phi_1) \,
    U_1^{'}(\Phi_1,r_1) \, \mathcal{P}_{1 \to 2}(\Phi_2)
    \nn \\
    &+\frac{\df \sigma^{\mathrm{LO}_{2}}}{\df \Phi_2}+\left[\frac{\df
        \sigma^{\mathrm{NLL}'_{r_1}}}{\df \Phi_1 \, \df r_1}
      -\frac{\df \sigma^{\mathrm{NLL}'_{r_1}}}{\df \Phi_1 \, \df
        r_1}\Bigg \vert_{\mathrm{LO}_2}\right] \mathcal{P}_{1 \to
      2}(\Phi_2) \Bigg \} \, \theta(r_1 >
    r_1^\mathrm{cut})\,\theta(r_0 > r_0^\mathrm{cut})
    \nn \\
    & + \frac{\df \sigma_\mathrm{nonproj}^{\mathrm{LO}_2}}{\df \Phi_2}
    \, \theta(r_1 < r_1^\mathrm{cut}) \, \theta(r_0 >
    r_0^\mathrm{cut}).
\end{align}
\end{widetext}

In the above equations, $\df \sigma^{\NtLL_{r_N},\mathrm{NLL}'_{r_N}}
/ \df \Phi_N(r_N)$ and $\df \sigma^{\NtLL_{r_N},\mathrm{NLL}'_{r_N}} /
\df \Phi_N \, \df r_N$ are respectively the cumulant and spectrum of
the $r_N$ N$^3$LL or NLL$'$ resummed cross section and the notation
$\vert_{\mathrm{(N)LO}_M}$ indicates their fixed-order expansion at
the (N)LO order of the process with $M$ final-state
partons. Furthermore, $U_1(\Phi_1, r_1)$ denotes the NLL $r_1$ Sudakov
form factor, $U_1'(\Phi_1, r_1)$ its derivative with respect to $r_1$,
and the functions $\mathcal{P}_{N \to N+1}(\Phi_{N+1})$ (called
splitting functions in the \geneva literature) are used to spread the
resummed $\Tau_N$ spectrum over the $\Phi_{N+1}$ phase space. They are
defined such that
\begin{align}
  \label{eq:splitting_functions}
  \int \frac{\df \Phi_{N+1}}{\df \Phi_N \, \df r_N} \, \mathcal{P}_{N
    \to N+1}(\Phi_{N+1}) = 1.
\end{align}
Finally, the `nonproj' label is used to indicate the $\Phi_{N+1}$
phase space points that do not have a $\Phi_N$ projection through the
$\tilde{\Phi}_N$ mapping.

The $\tilde{\Phi}_1$ mapping is constructed in such a way that it
preserves the value of $r_0$, i.e.
\begin{align}
  r_0(\tilde{\Phi}_1(\Phi_2))= r_0(\Phi_2),
\end{align} 
which is crucial to ensure that the $r_1$ resummation does not spoil
the resummed $r_0$ spectrum. Since we use $r_0=q_T$ in this work, we
require a $q_T$-preserving mapping in our NLO$_1$
calculation. Mappings which satisfy this criterion have been
previously used in \refscite{Alioli:2015toa,Alioli:2021qbf}. In this
work, we construct a new hybrid mapping by combining the ISR part of
the $\Tau_0-q_T$ preserving mapping from \refcite{Alioli:2015toa}
with the FSR part of the mapping used in \refcite{Gavardi:2023aco}
which preserves the transverse momentum of the jet,
$p_T^\mathrm{jet}$. This has better numerical performance compared to
the $q_T$ preserving mapping of \refcite{Alioli:2021qbf} while still
avoiding unphysical artefacts in exclusive distributions, which are
caused by the FSR part of the $\Tau_0-q_T$ mapping.

The integration measure introduced in \eq{splitting_functions} is
defined as
\begin{align}
  \frac{\df \Phi_{N+1}}{\df \Phi_N \, \df r_N} =& \df \Phi_{N+1} \,
  \delta[\Phi_N - \bar{\Phi}_N(\Phi_{N+1})]
  \nn \\
  & \times \delta[r_N - r_N(\Phi_{N+1})]
\end{align}
and indicates the integration over the $\Phi_{N+1}$ phase space points
whose value of the $N$-jet resolution variable is $r_N$ and which are
projected on $\Phi_N$ by the $\bar{\Phi}_N$ mapping (called splitting
mapping in the \geneva literature). The $\bar{\Phi}_N$ mapping does
not need to coincide with $\tilde{\Phi}_N$. Indeed, to simplify the
implementation of the $\mathcal{P}_{1 \to 2}$ functions, in this work
we employ a splitting mapping $\bar{\Phi}_N \neq \tilde{\Phi}_N$,
whose details are discussed in section~\ref{sec:splitting_mapping}.

\section{Resummation of transverse momentum in SCET}
\label{sec:scetqt}
The factorisation of the leading-power $q_T$ spectrum was first established in \refscite{Collins:1981uk,Collins:1981va, Collins:1984kg}, and extended in \refscite{Catani:2000vq, deFlorian:2001zd, Collins:2011zzd}. In this paper, we adopt the SCET framework \cite{Bauer:2000ew, Bauer:2000yr, Bauer:2001ct, Bauer:2001yt,Bauer:2002nz}, where $q_T$ factorisation was formulated in \refscite{Becher:2010tm, Echevarria:2011epo, Chiu:2012ir, Li:2016axz}. In particular, we employ rapidity renormalisation \cite{Chiu:2011qc, Chiu:2012ir} using the exponential regulator \cite{Li:2016axz}. In this formulation, the singular cross section for colour-singlet production can be written as 
\begin{align} \label{eq:tmd_factorization1}
\frac{\df \sigma^\sing}{\df q_T \, \df \Phi_0}
&= \sum_{a,b} H_{ab}(\Phi_0; \mu) \int \! \df^2 \vec{k}_a \, \df^2 \vec{k}_b \, \df^2 \vec{k}_s \,
\\\nn & \qquad \times
\delta(q_T - |\vec{k}_a - \vec{k}_b - \vec{k}_s|)
B_a(x_a, \vec{k}_a; \mu, \nu/\omega_a)
\\\nn & \qquad \times
B_b(x_b, \vec{k}_b; \mu, \nu/\omega_b)\,
S_{ab}(\vec{k}_s; \mu, \nu)
\, .
\end{align}
The hard function $H_{ab}$ is process-dependent and describes the hard
interaction producing the colour singlet $ab\to F$, where $a,b$ denote
the available partonic channels at leading order. The beam functions
$B_{a,b}$ describe collinear radiation with total transverse momentum
$\vec{k}_{a,b}$ and longitudinal momentum fractions $x_{a,b}= (m
/{\Ecm}) e^{\pm Y}$ (for colour-singlet mass $m$, rapidity $Y$ and
hadronic centre-of-mass energy $\Ecm$), with $\omega_{a,b} = x_{a,b}
\Ecm$. The soft function $S_{ab}$ describes soft radiation with total
transverse momentum $\vec{k}_s$. Finally, the scales $\mu$ and $\nu$
denote the virtuality and rapidity renormalisation scales.

To achieve all-order resummation, each of the functions in \eq{tmd_factorization1} is first evaluated at its natural scale(s) $\mu_{H,B,S}$, $\nu_{B,S}$, and then evolved to a common set of scales $\mu$, $\nu$ by solving a coupled system of renormalisation group equations. As shown in \refcite{Ebert:2016gcn}, the exact solution of said equations in $q_T$ space in terms of distributions is equivalent to the canonical solution in $b_T$ space appropriately transformed. In this work we follow common practice and employ the latter solution. For details see \refscite{Ebert:2020dfc, Ebert:2016gcn, Billis:2019evv}.

While the resummed cross section in \eq{tmd_factorization1} is appropriate when describing events at low $q_T$, for higher values it is no longer adequate and we must rely on a calculation at fixed order in perturbation theory. The matching of resummed and fixed order calculations is achieved additively as detailed in sec.~\ref{sec:geneva} -- we are still required, however, to specify a prescription which switches off the resummation at large $q_T$ to ensure a smooth transition between the resummed and fixed-order parts of the calculation. This is achieved using hybrid profile scales~\cite{Ligeti:2008ac, Abbate:2010xh,Lustermans:2019plv}, which allow $\mu_H,\mu_B,\mu_S$ to take different values in the resummation region (required to minimise the size of large logarithms), but which flow to a common scale with increasing $q_T$, thus naturally ending RG evolution for $q_T\sim Q$.

The framework described above is sufficiently general to describe the production of any colour-singlet final state (given an appropriate hard function). However, implementation details are in general process-dependent. In particular, the shape of the profile scales which determine the transition between resummation and fixed-order regions of phase space is normally chosen based on the relative size of the singular and nonsingular contributions as a function of $q_T$ and differs between e.g.~Drell-Yan and Higgs boson production. In our specific case of Higgs boson production in heavy-quark annihilation, these considerations were examined in \refcite{Cal:2023mib}. Since we directly interface the \scetlib $q_T$ module~\cite{Billis:2024dqq} used in that work to \geneva, we employ the same central scale choices and profile shapes as used there. We also estimate our uncertainties via profile scale variation in the same manner. We note that just as in \refcite{Alioli:2023har}, this allows us to separate dependence on the beam and factorisation scales $\mu_B$ and $\mu_F$, allowing us to perform 7-point scale variations for inclusive observables. For the sake of completeness, we detail our profile scale choices, transition points and variation prescription in \app{profilescales}.

\section{Resummation of one-jettiness with generalised measures}
\label{sec:tau1pt}
\subsection{Observable definition}
The $N$-jettiness observable is most often defined as
\begin{align}
\Tau_N = \sum_i\min_m\left\{\frac{2q_m\cdot p_i}{Q_m}\right\}\,,
\label{eq:TauNinvmass}
\end{align}
where the sum runs over the four-momenta of all coloured particles $p_i^\mu$, the minimisation over $m$ runs over all beam and jet reference momenta $q_m^\mu$, and the factors $Q_m$ are normalisation factors. This definition uses an invariant mass-like measure and obeys a SCET-I type factorisation, which has been well-studied in the literature. 

It was remarked in the original $N$-jettiness paper, however, that the definition in \eq{TauNinvmass} can be extended to use generic measures in each beam or jet region~\cite{Stewart:2010tn}. Moreover, $N$-jettiness can be thought of not merely as an event-shape, but as a way to define an exclusive jet algorithm which partitions the phase space into two beam and $N$ jet regions, to which emissions are assigned. The measure factors which are used to assign emissions to each region need not be identical to the value which the observable returns. In general, therefore, we must define separate distance measures $d_m(p_i)$ which assign the emission $i$ to one of the $N+2$ regions $m$ if 
\begin{align}
  \label{eq:distance}
  d_m(p_i) = \min\big\{d_1(p_i),\dots,d_N(p_i), d_a(p_i),d_b(p_i)\big\}\,,
\end{align}
and the resulting observable $f_m$ which is returned by that emission once it falls into the region $m$,
\begin{align}
\Tau^{(m)}=\sum_{i\in m} f_m(\eta_i,\phi_i)p_{T,i}\,,
\end{align}
where $\eta_i$, $\phi_i$ and $p_{T,i}$ denote the pseudorapidity, azimuth and transverse momentum associated with $i$. We remark that the definitions of the jet regions $d_j$ for $j\in \{1,\dots,N\}$ are themselves dependent on a determination of the jet axes, which may be performed either by running an exclusive jet algorithm (e.g. anti-$k_T$) over the final state, or by a minimisation of the $N$-jettiness over all possible axes.
The latter definition has the advantage of guaranteeing insensitivity to soft-recoil effects, whose correct treatment requires the introduction of additional transverse momentum convolutions in the factorisation formula. At one-loop, it is equivalent to using the winner-take-all (WTA) axis~\cite{Larkoski:2014uqa}, in that the jet direction is determined solely by the `hardest' emission. 

This generalised definition of the $N$-jettiness has also been well-studied in the literature, and the choice of the measures has important consequences for the structure of the factorisation. The XCone algorithm implements precisely the procedure described above to act as an exclusive jet algorithm, and in that context a number of different measure choices have been studied~\cite{Stewart:2015waa}. \Refcite{Bertolini:2017efs} completed the study of factorisation types (i.e., all possible combinations of SCET-I/SCET-II measures for beams and jets) and provided the one-loop soft functions relevant for colour singlet production in association with a single jet. In addition, the $N$-subjettiness observable, introduced in \refcite{Thaler:2010tr} as a way to study jet substructure, uses choices of the $f_m$ which are transverse momentum-like by default. 

In this work, we will make specific choices for both the $d_m$ and $f_m$ and use $N=1$, thus defining a generalised one-jettiness observable which we refer to as $\Tau_1^{p_T}$. Specifically, we choose the conical measure~\cite{Stewart:2015waa,Thaler:2011gf} for the region assignment, which (for isolated jets) clusters in a manner equivalent to the anti-$k_T$ algorithm:
\begin{align}
  \label{eq:Tau1pT_distance}
d_0(p_i)=1,\qquad d_{m\geq 1}(p_i)=\frac{R_{im}^2}{R^2}.
\end{align}
In the above equation, we have defined a single beam measure,
\begin{align}
d_0(p_i)=\min\{d_a(p_i),d_b(p_i)\}\,,
\end{align}
and introduced the distance
\begin{align}
R_{im}\equiv \sqrt{(\eta_i-\eta_m)^2+(\phi_i-\phi_m)^2}
\end{align}
and (constant) jet radius $R$. For the measurement itself, defining
\begin{align}
\label{eq:TaupT}
\Tau_1^{p_T}=\sum_i\begin{cases}
p_{T,i}\,f_B(\eta_i), & \text{for } d_B(p_i) <d_J(p_i),  \\
p_{T,i}\,f_J(\eta_i,\phi_i), & \text{for } d_J(p_i) <d_B(p_i), 
\end{cases}
\end{align}
we choose the boost-invariant generalisation of broadening~\cite{Banfi:2004nk,Bertolini:2017efs} for the jet region,
\begin{align}
f_J(\eta_i,\phi_i) &= \sqrt{2\cosh(\eta_i-\eta_J)-2\cos(\phi_i-\phi_J)} \nn \\
&\equiv \mathcal{R}_{iJ} 
\end{align}
where the distance $d_J(p_i)$ is measured with respect to the jet axis
$J$ according to \eq{Tau1pT_distance}, and the transverse energy for
the beam region,
\begin{align}
f_B(\eta_i)=1\,,
\end{align}
with the distance $d_B(p_i)$ being equal to $d_0(p_i)=1$. Both
observables are of SCET-II type -- the setup described above
corresponds to one of the choices made in \refcite{Bertolini:2017efs},
for which the one-loop soft function was calculated. It was shown in
that work that the cross section then factorises as
\begin{align}
\frac{\mathrm{d}\sigma_\kappa}{\mathrm{d}\Phi_1 \, \mathrm{d}\TaupT} &=
H_\kappa(\Phi_1,\mu) \int \left(\prod_n \mathrm{d}k_n\right)\nn \\
& \times S_\kappa\left(\TaupT-\sum_i k_i,\{n_m\},\{d_m\},\mu,\frac{\nu}{\mu}\right) \nn \\
& \times B_{\kappa_a}\left(k_a,x_a,\mu,\frac{\nu}{\omega_a}\right)B_{\kappa_b}\left(k_b,x_b,\mu,\frac{\nu}{\omega_b}\right)\nn \\
&\times J_{\kappa_j}\left(k_j,\mu,\frac{\nu}{\omega_j}\right)\,,
\label{eq:taufac}
\end{align}
where the channel index $\kappa \equiv \{\kappa_a,\kappa_b,\kappa_j\}$ runs over all possible flavours of incoming and outgoing partons. 

\subsection{Ingredients and resummation at NLL$'$}
At NLL$'$ accuracy, each of the hard, soft, beam and jet functions are required at one-loop accuracy as are the noncusp anomalous dimensions, while the cusp anomalous dimension is needed one order higher. The hard function is easily obtained from the corresponding one-loop QCD amplitude.\footnote{Although in general both hard and soft functions are matrices in colour space, for the case of $N=1$ the colour algebra diagonalises and it is possible to rewrite the factorisation in the form \eq{taufac}.} It satisfies an evolution equation
\begin{align}
\mu \frac{\mathrm{d}}{\mathrm{d}\mu} \ln H_{\kappa}(\Phi_1, \mu) = \gamma_H^{\kappa}(\Phi_1,\mu)
\end{align}
where the hard anomalous dimension is of the form
\begin{align}
\gamma^\kappa_H(\Phi_1,\mu) &= \Gamma_\cusp[\as(\mu)] \biggl[ \mathbf{T}_a^2 \ln\frac{\omega_a^2e^{-2\eta_J}}{\mu^2}\nn \\
    &\qquad+ \mathbf{T}_b^2 \ln\frac{\omega_b^2e^{2\eta_J}}{\mu^2} + \mathbf{T}_j^2 \ln\frac{\omega_j^2}{(2\cosh\eta_J)^2\mu^2} \biggr]\nn\\
    &\quad + \gamma^\kappa_H[\as(\mu)]
\,, \\
\gamma^\kappa_H[\as(\mu)] &=  (n_q + n_{\bar{q}}) \gamma^q[\as(\mu)] + n_g \gamma^g[\as(\mu)]
\label{eq:hard_noncusp}
\,,\end{align}
where $n_q$ and $n_g$ label respectively the number of quarks and gluons in the channel $\kappa$.

The relevant soft function was calculated in \refcite{Bertolini:2017efs}. Its evolution equation reads
\begin{align}
\mu \frac{\mathrm{d}}{\mathrm{d}\mu} S_{\kappa}&\left(k_S, \Phi_1, \mu,\frac{\nu}{\mu}\right) = \nn \\& \int \mathrm{d}k_S'\gamma_{\mu,S}^{\kappa}\left(k_S-k_S',\Phi_1,\mu,\frac{\nu}{\mu}\right)S_{\kappa}\left(k_S', \Phi_1, \mu,\frac{\nu}{\mu}\right)
\end{align}
where we have suppressed the dependence on the $d_m$ for brevity and translated the dependence on the reference directions $n_m$ into a dependence on the phase space $\Phi_1$. The soft virtuality anomalous dimension is given to all orders by 
\begin{align}
\gamma_{\mu,S}^{\kappa}\left(k_S,\Phi_1,\mu,\frac{\nu}{\mu}\right) &= 2\Gamma_\cusp[\as(\mu)]\,\delta(k_S)\nn \\
&\times \biggl[ -\left(\mathbf{T}_a^2+\mathbf{T}_b^2+\mathbf{T}_j^2\right)\ln\frac{\nu}{\mu} \nn \\
    &\quad+ \mathbf{T}_j^2\ln(2\cosh\eta_J) + (\mathbf{T}_a^2-\mathbf{T}_b^2) \eta_J\biggr]\nn\\
    &\quad + \gamma^\kappa_{\mu,S}[\as(\mu)]\,\delta(k_S)
\end{align}
where the noncusp term \mbox{$\gamma^{\kappa\,(0)}_{\mu,S}[\as(\mu)]=0$} at one-loop.
The soft function also obeys a rapidity evolution equation
\begin{align}
\nu \frac{\mathrm{d}}{\mathrm{d}\nu} S_{\kappa}&\left(k_S, \Phi_1, \mu,\frac{\nu}{\mu}\right) = \nn \\& \int \mathrm{d}k_S'\gamma_{\nu,S}^{\kappa}\left(k_S-k_S',\mu\right)S_{\kappa}\left(k_S', \Phi_1, \mu,\frac{\nu}{\mu}\right)
\end{align}
where the rapidity anomalous dimension at one-loop order is given by 
\begin{align}
\gamma_{\nu,S}^{\kappa\,(1)}(k_S,\mu)&= 2\Gamma_0\frac{\as(\mu)}{4\pi} \biggl[ \left(\mathbf{T}_a^2+\mathbf{T}_b^2+\mathbf{T}_j^2\right)\frac{1}{\mu}\mathcal{L}_0\left(\frac{k_S}{\mu}\right)\biggr]
\end{align}
and the plus distribution $\mathcal{L}_0(x)$ is defined as in \app{plusdistrib}.

The collinear beam functions relevant for the observable are those calculated at one-loop in e.g. \refscite{Becher:2012qa,Becher:2013xia,Stewart:2013faa} while the recoil-free broadening jet function was calculated in \refcite{Larkoski:2014uqa}. Both beam and jet functions obey a set of similar evolution equations,
\begin{align}
\mu \frac{\mathrm{d}}{\mathrm{d}\mu} B_{i}&\left(k_B, x, \mu,\frac{\nu}{\omega}\right) = \nn \\& \int \mathrm{d}k_B'\gamma_{\mu,B}^{i}\left(k_B-k_B',\mu,\frac{\nu}{\omega}\right)B_{i}\left(k_B', x, \mu,\frac{\nu}{\omega}\right) \,,\\
\nu \frac{\mathrm{d}}{\mathrm{d}\nu} B_{i}&\left(k_B, x, \mu,\frac{\nu}{\omega}\right) = \nn \\& \int \mathrm{d}k_B'\gamma_{\nu,B}^{i}\left(k_B-k_B',\mu\right)B_{i}\left(k_B', x, \mu,\frac{\nu}{\omega}\right)\,, \\
\mu \frac{\mathrm{d}}{\mathrm{d}\mu} J_{i}&\left(k_J, \mu,\frac{\nu}{\omega}\right) = \nn \\& \int \mathrm{d}k_J'\gamma_{\mu,J}^{i}\left(k_J-k_J',\mu,\frac{\nu}{\omega}\right)J_{i}\left(k_J', \mu,\frac{\nu}{\omega}\right) \,,\\
\nu \frac{\mathrm{d}}{\mathrm{d}\nu} J_{i}&\left(k_J, \mu,\frac{\nu}{\omega}\right) = \nn \\& \int \mathrm{d}k_J'\gamma_{\nu,J}^{i}\left(k_J-k_J',\mu\right)J_{i}\left(k_J', \mu,\frac{\nu}{\omega}\right)\,,
\end{align}
where the virtuality anomalous dimensions take the forms
\begin{align}
\gamma_{\mu,B}^i\left(k_B,\mu,\frac{\nu}{\omega}\right)&=\left\{2\mathbf{T}_i^2\Gamma_{\mathrm{cusp}}\log\frac{\nu}{\omega}+\gamma_B^i[\alpha_s(\mu)]\right\}\delta(k_B)\,, \\
\gamma_{\mu,J}^i\left(k_J,\mu,\frac{\nu}{\omega}\right)&=\left\{2\mathbf{T}_i^2\Gamma_{\mathrm{cusp}}\log\frac{\nu}{\omega}+\gamma_J^i[\alpha_s(\mu)]\right\}\delta(k_J)\,,
\end{align}
and at one loop $\gamma^{i\,(0)}_{B/J}[\alpha_s(\mu)]=-2\gamma^{i\,(1)}[\as(\mu)]$. The one-loop rapidity anomalous dimensions are instead given by
\begin{align}
\gamma_{\nu,B}^i(k_B,\mu)&= -2\Gamma_0\frac{\as(\mu)}{4\pi}\mathbf{T}_i^2\frac{1}{\mu}\mathcal{L}_0\left(\frac{k_B}{\mu}\right)\\
\gamma_{\nu,J}^i(k_J,\mu)&= -2\Gamma_0\frac{\as(\mu)}{4\pi}\mathbf{T}_i^2\frac{1}{\mu}\mathcal{L}_0\left(\frac{k_J}{\mu}\right)\,.
\end{align}
The evolution equations in $\mu$ and $\nu$ are most easily solved by transforming to Laplace space, first evolving from $\nu_X\to \nu$ at fixed $\mu_X$ and subsequently from $\mu_X\to \mu$ (for $X\in \{B,S,J\}$). At NLL, the solution for the soft function can be written analytically as
\begin{widetext}
\begin{align}
S_{\kappa}&\left(k_S, \Phi_1, \mu,\frac{\nu}{\mu}\right) =\nn \\
&\exp\Bigg\{2(C_{\kappa_a}+C_{\kappa_b}+C_{\kappa_j})K_\Gamma(\mu_S,\mu)+\Big[2(C_{\kappa_a}-C_{\kappa_b})\,\eta_J+2C_{\kappa_j}\ln(2\cosh\eta_J)+2(C_{\kappa_a}+C_{\kappa_b}+C_{\kappa_j})\ln\frac{\mu_S}{\nu}\Big]\eta_\Gamma(\mu_S,\mu)\Bigg\} \nn \\
&\times\int \mathrm{d}k_S' 
\mathcal{V}_{\eta_S^\kappa}(k_S-k_S',\mu_S)\,
S_{\kappa}\left(k_S', \Phi_1, \mu_S,\frac{\nu_S}{\mu_S}\right)\,,
\end{align}
\begin{align}
\eta_S^\kappa&=\frac{\as(\mu_S)}{4\pi}2\Gamma_0\,(C_{\kappa_a}+C_{\kappa_b}+C_{\kappa_j})\ln\left(\frac{\nu}{\nu_S}\right)\,,
\end{align}

where the function $\mathcal{V}_a(k,\mu)$ is defined in \app{plusdistrib} and the quadratic Casimir factors $C_i=\mathbf{T}_i^2$. Similarly, the solution for the beam and jet functions can be written as 
\begin{align}
B_{i}&\left(k_B, x, \mu,\frac{\nu}{\omega}\right) =\nn \\
&\exp\Bigg\{K_\gamma(\mu_B,\mu)+2C_{i} \ln\frac{\nu}{\omega}\,\eta_\Gamma(\mu_B,\mu)\Bigg\} \int \mathrm{d}k_B' 
\mathcal{V}_{\eta_B^i}(k_B-k_B',\mu_B)\,
B_{i}\left(k_B', x, \mu_B,\frac{\nu_B}{\mu_B}\right)\,,\\
&\qquad\qquad\eta_B^i=-\frac{\as(\mu_B)}{4\pi}2\Gamma_0\, C_{i}\ln\left(\frac{\nu}{\nu_B}\right)\,,\\
J_{i}&\left(k_J, \mu,\frac{\nu}{\omega}\right) =\nn \\
&\exp\Bigg\{K_\gamma(\mu_J,\mu)+2C_{i} \ln\frac{\nu}{\omega}\,\eta_\Gamma(\mu_J,\mu)\Bigg\} \int \mathrm{d}k_J' 
\mathcal{V}_{\eta_J^i}(k_J-k_J',\mu_J)\,
J_{i}\left(k_J', \mu_J,\frac{\nu_J}{\mu_J}\right)\,,\\
&\qquad\qquad\eta_J^i=-\frac{\as(\mu_J)}{4\pi}2\Gamma_0\, C_{i}\ln\left(\frac{\nu}{\nu_J}\right)\,.
\end{align}

\begin{table}
    \centering
    \renewcommand{\tabcolsep}{2ex}
    \begin{tabular}{Sl |c c c c}
    \hline \hline
         ${\sigma^{\mathrm{NNLO}}_{b \bar{b}\to H} \hspace{2ex}[\mathrm{fb}]}$& SusHi &
         $q_T^{\mathrm{cut}}= 1\GeV$ &$q_T^{\mathrm{cut}}= 5\GeV$ \, &
         $q_T^{\mathrm{cut}}= 10\GeV$\,\\
         \hline
         $\mu =m_H$& $543.5 \pm 0.5$ &$544.3 \pm 1.0$&$544.9\pm 0.8$&$548.6 \pm 0.4$\\
         $\mu =m_H/2$ &$518.0 \pm0.5$&$518.5 \pm 1.1$&$517.8 \pm 0.8$& $518.4 \pm 0.5 $\\
         $\mu =2m_H$ &$581.7\pm 0.6$ &$583.3 \pm 0.9$& $585.1 \pm 0.7$& $591.4 \pm 0.4$\\
    \hline \hline
    \end{tabular}
    \caption{Comparison between \textsc{SusHi} and \geneva of predictions for the NNLO total cross section in $b\bar{b}\to H$. \geneva predictions at various values of $q_T^\cut$ are shown, and at various values of $\mu=\mu_R=\mu_F$.}
    \label{tab:qTcutsigmas}
\end{table}
\end{widetext}
For convenience, we reproduce the one-loop expressions for soft, beam and jet functions in \app{functions}. We note that while it is possible to write down the solutions of the evolution equations analytically at NLL (NLL$'$) order, extending the above solutions to higher resummed orders seems challenging. The underlying reason for this is a combination of the facts that the evolution equations are solved in Laplace space and that  the rapidity anomalous dimensions have a non-trivial dependence on $k_{B,S,J}$. At one-loop order, this necessitates taking an inverse Laplace transform of functions of the form $\exp(-a \ln x)$, for constant $a$. While this is still possible analytically at this order, the appearance of higher powers of the logarithm in the exponent (arising ultimately from higher plus distributions $\mathcal{L}_{1,2,\dots}$ in $\gamma_\nu$) hinders an analytic treatment starting at NNLL. Resummation at this order is therefore likely to require a numerical implementation of the inverse integral transform, as is normally performed for $q_T$ resummation. 

\subsection{Splitting mappings}
\label{sec:splitting_mapping}

To generate hard events that are distributed according to the
NLL$^\prime$ $\TaupT$ resummed spectrum, \geneva assigns the events a
weight obtained by multiplying the spectrum by a function (called
splitting function in the \geneva literature) of the
$\mathrm{d}\Phi_2$ phase space that integrates to 1 for every
underlying $\Phi_1$ phase space point and $\TaupT$ value:
\begin{equation}
  \int \frac{\mathrm{d}\Phi_2}{\mathrm{d}\Phi_1 \> \mathrm{d}\TaupT} \> \mathcal{P}_{1 \to 2}\left(\Phi_2\right)
  = 1.
\end{equation}
The details of the numeric implementation and the functional forms of
the splitting functions used by \geneva are provided in
\refscite{Alioli:2023har, Gavardi:2023oho}.

This procedure requires introducing a projection $\bar{\Phi}_1$ (called a splitting mapping in the \geneva literature) from the phase space $\mathrm{d}\Phi_2$ to the underlying phase space $\mathrm{d}\Phi_1$. The choice of such a projection is constrained by the requirement of infrared safety that univocally fixes the splitting mapping in the limit of small $\TaupT$. At large $\TaupT$, instead, the choice is largely arbitrary, to the point that not all the phase space points are even required to be projectable (a non-projectable $\Phi_2$ phase space point would simply imply that $P_{1 \to 2}(\Phi_2) = 0$). The arbitrariness of this choice might appear to translate into a large theoretical
uncertainty in the final distributions. This, however, is not the
case, since the splitting functions always multiply differences
between resummed spectra and their truncated perturbative expansion,
which, at large values of $\TaupT$, are beyond the accuracy of the
calculation. In other words, different choices of splitting mappings
introduce distortions in the distributions of the generated events,
whose size is formally subleading with respect to the claimed NNLO
accuracy.

For some distributions of particular importance, however, it might be
preferable to avoid these kinds of effects entirely. To that purpose,
we can exploit the freedom in the choice of the mapping to enforce the
preservation of a set of observables $O_i$, for which
\begin{equation}
  O_i\!\left(\Phi_2\right) = O_i\!\left(\bar{\Phi}_1\!\left(\Phi_2\right)\right).
\label{eq:1to2_qT_splitting_function_constraint}
\end{equation}
For these observables, the $\TaupT$ resummation can be implemented
in a unitary fashion, such that their distributions remain unaffected.

In this work, we require the splitting mapping to preserve the entire
colour singlet four-momentum. This guarantees that the NNLO accuracy
of the Higgs rapidity distribution and the N$^3$LL$^\prime$ accuracy
of the Higgs transverse momentum distribution are not spoiled. This
condition still leaves one degree of freedom that we fix with the
further constraint that the splitting mapping also preserves the
rapidity of the `hardest' final-state parton (i.e.,~the one that
defines the jet direction in the WTA clustering). The choice of
$\TaupT$ as 1-jet resolution variable does not constrain the mapping,
but affects the numeric implementation of the splitting
functions. Guaranteeing that
\eq{1to2_qT_splitting_function_constraint} is satisfied indeed
requires an analytic computation of the integration limits on $\Phi_2$
at fixed $\Phi_1 = \bar{\Phi}_1(\Phi_2)$ and $\Tau_1^{p_T}$, as well
as the measure $\mathrm{d}\Phi_2 / \mathrm{d}\Phi_1 \>
\mathrm{d}\Tau_1^{p_T}$.

\section{Results}
\label{sec:results}
In this section, we present our numerical results for the $q\bar q \to
H$ process in a proton-proton scattering. We use the following input
parameters: $\Ecm = 13\TeV$, Higgs boson mass $m_H=125\GeV$ and strong
coupling $\alpha_s(m_Z) = 0.118$ with $m_Z = 91.1876 \GeV$. Unless
otherwise indicated we use the \texttt{MSHT20nnlo} PDF
set~\cite{Bailey:2020ooq}. The Yukawa couplings are computed by
evolving the $\overline{\mbox{MS}}$ quark masses $\overline{m}_q(m_q)$
to the desired scale. The input values are
$\overline{m}_b(\overline{m}_b) =4.18\GeV$ and
$\overline{m}_c(\overline m_c) =1.27\GeV$
\cite{ParticleDataGroup:2022pth}. Finally, we set the Higgs vacuum
expectation value to $v = 246.22\GeV$.

\begin{figure}[t!]
    \centering
    \includegraphics[width=\linewidth]{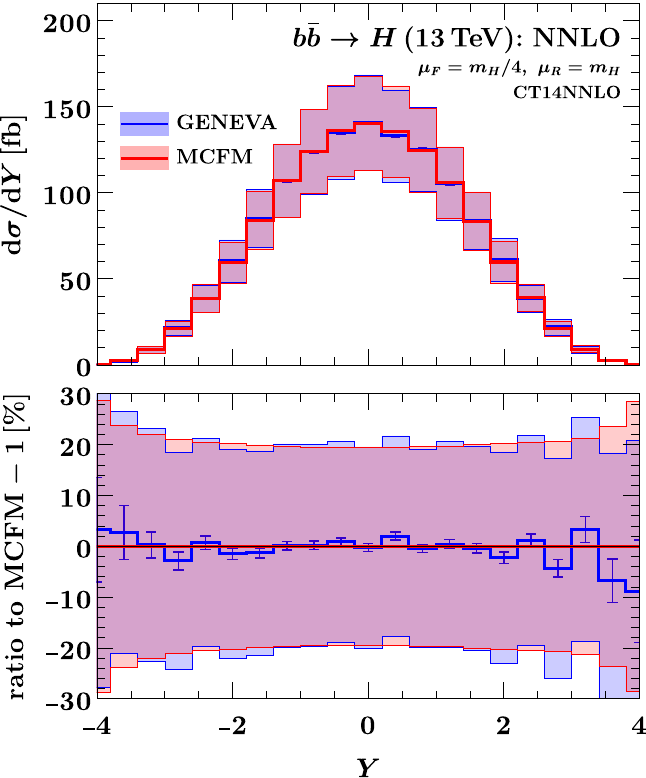}
    \caption{Comparison between MCFM and \geneva for the NNLO Higgs boson rapidity spectrum in $b\bar{b}\to H$. Uncertainty bands are obtained via a 7-pt. scale variation procedure.}
    \label{fig:MCFMval}
\end{figure}

\begin{figure}[t!]
    \centering
    \includegraphics[width=\linewidth]{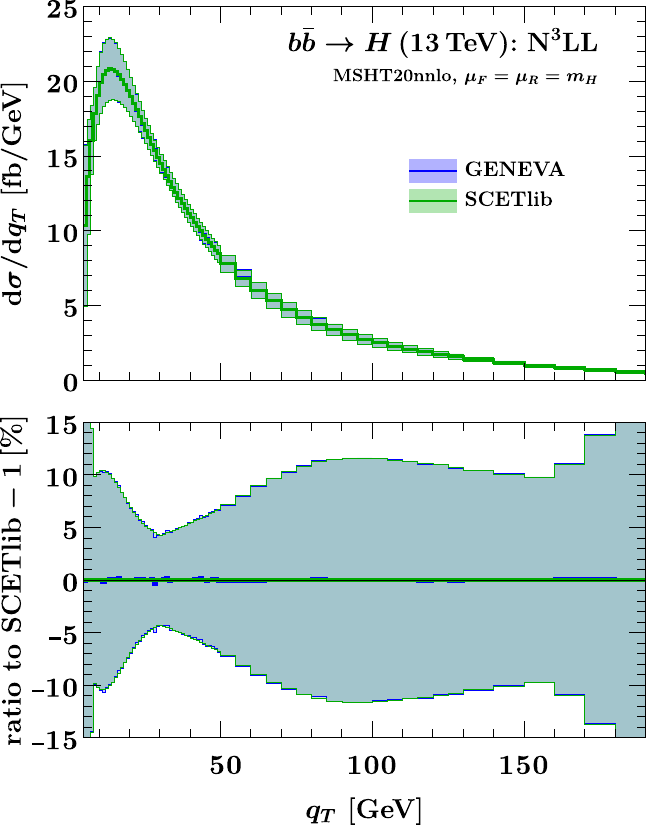}
    \caption{Comparison between \scetlib and \geneva for the N$^3$LL Higgs boson $q_T$ spectrum in $b\bar{b}\to H$. Uncertainty bands are obtained as explained in \app{profilescales}}
    \label{fig:SCETlibval}
\end{figure}

\subsection{Fixed order validation}
\label{sec:FOval}

\begin{figure*}[ht!]
	\centering
	\begin{minipage}[t]{0.48\textwidth}
		\centering
		\centering
		\includegraphics[width=\textwidth]{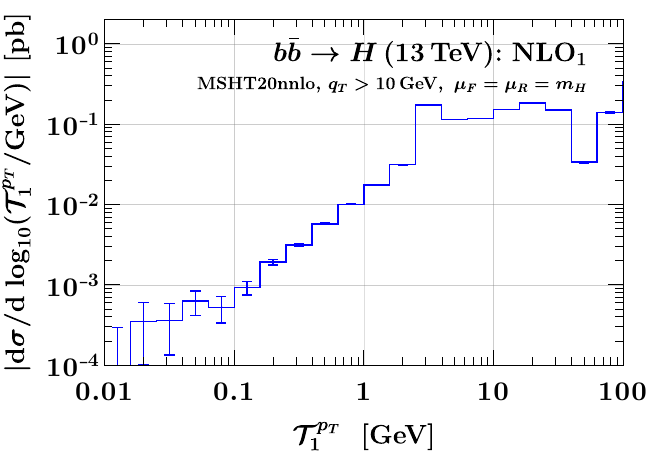}
		\caption{Nonsingular $\TaupT$ spectrum for $b\bar{b}\to H$.}
		\label{fig:tau1nons}
	\end{minipage}
	\begin{minipage}[t]{0.48\textwidth}
		\centering
		\includegraphics[width=\textwidth]{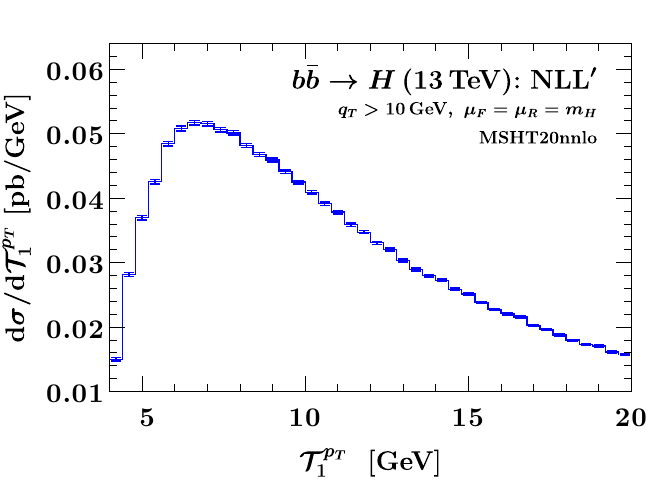}
		\caption{Resummed $\TaupT$ spectrum for $b\bar{b}\to H$.}
		\label{fig:tau1res}
	\end{minipage}
\end{figure*}

\begin{figure}[t!]
	\centering
	\includegraphics[width=\linewidth]{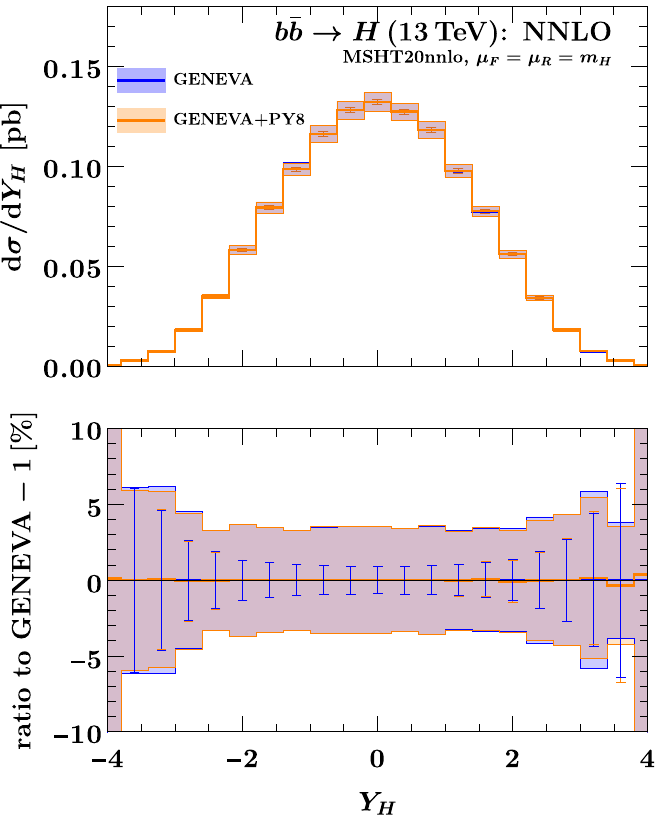}
	\caption{Higgs boson rapidity spectrum in $b\bar{b}\to H$. Results are shown at parton level and after showering.}
	\label{fig:bbHyH}
\end{figure}
We begin by validating our results at fixed order in perturbation theory. Before presenting our numerical comparisons, we take this opportunity to remind the reader of a subtlety in choosing the renormalisation scale in a resummed calculation, which will affect our results. The profile scales which we employ in \geneva are functions of the resolution variable $q_T$. The consequence of this is that the operations of setting the scale and integrating the $q_T$ spectrum to obtain the total cross section (or an inclusive distribution such as the colour singlet rapidity) do not commute. This leaves some freedom in how we present our results -- we can either set scales in the spectrum and obtain inclusive quantities by integration, or instead set scales in the integrated cross section and obtain the spectrum by differentiation. The two options differ by formally higher-order terms, which however can be numerically sizeable. The first choice is more natural when one is interested in the $q_T$ spectrum, since it matches what one would normally do in a resummed calculation. The latter instead guarantees that the results of a fixed order calculation (say, the NNLO cross section) are recovered, but may result in distortions of the $q_T$ spectrum. 

In \refcite{Alioli:2015toa}, a compromise was proposed. The key idea is to set scales in the spectrum, but to add back a higher-order term which is given by the difference of the spectrum and the $q_T$-derivative of the cumulant. This is further modulated by an interpolating function $\kappa(q_T)$, which limits the activity of the term to the small $q_T$ region. In this way, one obtains the correct fixed order normalisation for inclusive quantities while minimising any distortions of the $q_T$ spectrum. In this work, we have chosen not to adopt this approach. For the $b\bar{b}\to H$ process in particular, the difference between spectrum and cumulant scale setting is a numerically large effect (in contrast to e.g. Drell-Yan, see \app{cumulant_scales}), and our predictions would be extremely sensitive to the exact way in which the cross section fix was implemented. In the rest of this subsection, we have set scales in the cumulant (which facilitates the comparison of our results with fixed order codes). In following subsections, however (and in particular when examining exclusive distributions such as the $q_T$ spectrum), we have instead adopted spectrum scale setting. This is in general more appropriate for Higgs boson production in heavy-quark fusion, since measurements of the Higgs boson $q_T$ spectrum are likely to be one of the few ways in which this process can be probed -- spectrum scale setting will therefore lead to the `best' description for this observable.

\begin{figure*}
	\centering
	\begin{minipage}[t]{0.45\textwidth}
		\centering
		\includegraphics[width=\textwidth]{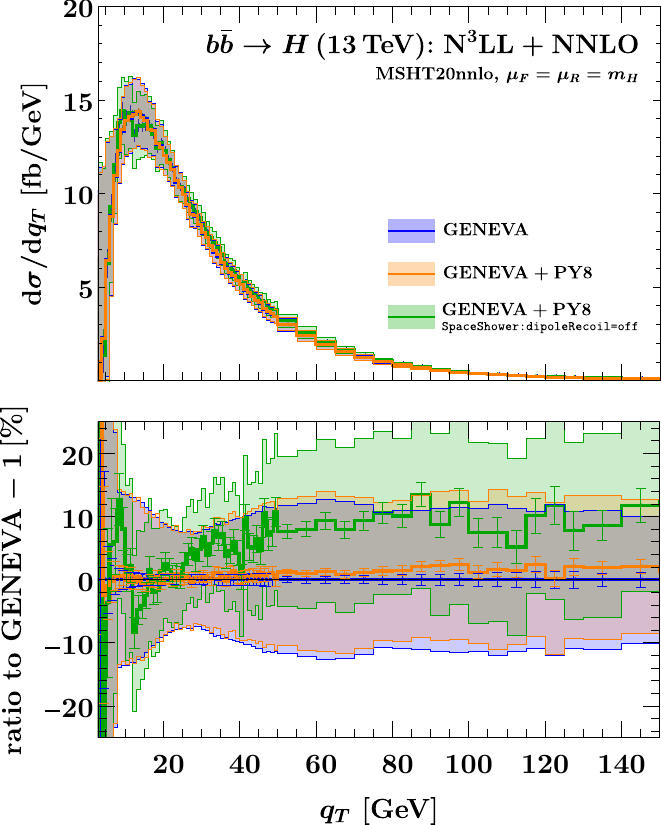}
		\caption{Higgs boson $q_T$ spectrum in $b\bar{b}\to H$. Results are shown at parton level and after showering.}
		\label{fig:bbHqT}
	\end{minipage}
	\hfill
	\begin{minipage}[t]{0.45\textwidth}
		\centering
		\includegraphics[width=\textwidth]{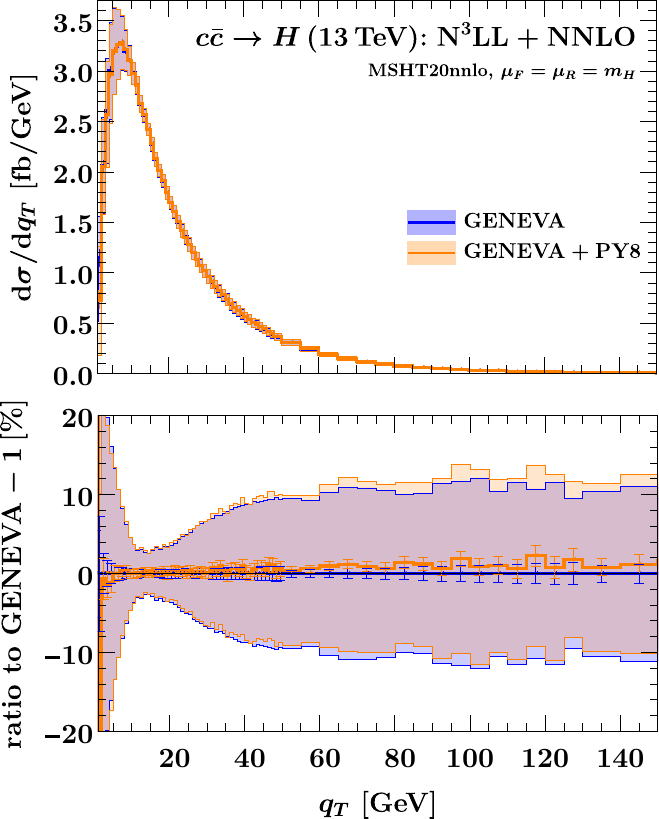}
		\caption{As in fig.\ref{fig:bbHqT}, but for $c\bar{c}\to H$.}
		\label{fig:ccHqT}
	\end{minipage}
\end{figure*}
The \geneva predictions for the inclusive cross section at various values of $q_T^\cut$ are shown in tab.~\ref{tab:qTcutsigmas}, and compared to NNLO predictions obtained from \textsc{SusHi}~\cite{Harlander:2012pb}. The absence of a fully local NNLO subtraction means that the implementation of \eq{0-jet_cross_section} in \geneva is done only in an approximate fashion, dropping all fixed order NNLO terms and relying on the singular approximation to the cross section at leading power to provide the two-loop information~\cite{Alioli:2015toa, Alioli:2019qzz}. This is exactly equivalent to an NNLO calculation in the limit $q_T^\cut \to 0$; however, for finite values one misses $\mathcal{O}(\alpha_s^2)$ power corrections below $q_T^\cut$. It is therefore desirable to set the cut as small as possible to minimise the effect of these inclusive nonsingular terms. The differences we observe with respect to the exact calculation implemented in \textsc{SusHi} are minor, reaching only 1\% even at very large values $q_T^\cut=10~\mathrm{GeV}$: moreover, for $q_T^\cut=1~\mathrm{GeV}$ the \geneva cross section is equivalent to that returned by \textsc{SusHi} to within the statistical precision of the calculation ($\sim0.2\%$). In the following we therefore will use $q_T^\cut=1~\mathrm{GeV}$ as our default choice. We also remark that, as in several previous implementations of \geneva for colour singlet production~\cite{Alioli:2021qbf,Alioli:2023har,Gavardi:2023aco}, we do not need to employ a reweighting procedure to account for missing nonsingular terms below $r^\cut$. This need has in any case been recently obviated by the development of improved subtraction techniques in \geneva, which combine the projection-to-Born method with slicing to allow very small values of $r^\cut$ to be reached while correctly accounting for fiducial power corrections~\cite{Alioli:2025hpa}.

In fig.~\ref{fig:MCFMval} we validate our predictions for the NNLO Higgs boson rapidity spectrum against an independent calculation from MCFM, obtained using zero-jettiness slicing~\cite{Boughezal:2016wmq,Mondini:2021nck}. Note that compatibility with MCFM requires us to set in this case (and in this case alone) $\mu_F=m_H/4,~\mu_R=m_H$ and to use the \texttt{CT14NNLO} PDF set~\cite{Dulat:2015mca}. We observe excellent agreement between the two calculations, both for the central value and for the 7-point scale variation bands.

\subsection{Resummed validation}

The \geneva method relies on an additive matching procedure of resummed and fixed-order calculations. It should therefore be the case that predictions for the $0-$jet resolution variable (in this case, $q_T$) from the event generator are exactly N$^3$LL accurate by construction, at least at parton level. In fig.~\ref{fig:SCETlibval} we confirm this explicitly by comparison with a separate run of \textsc{SCETlib}. We observe complete agreement for both central value and uncertainty bands, thus corroborating the correctness of our implementation.

\subsection{The $1/2-$jet bin separation}
A novel feature of this work is the use of $\TaupT$ as a resolution variable to separate the $1-$ and $2-$jet bins. We remind the reader that, in addition to the resummation of this variable at NLL$'$, the implementation of this variable in the event generator has required us to develop appropriate mappings for the splitting functions which spread the resummation over the higher multiplicity phase space. Both issues are detailed in sec.~\ref{sec:tau1pt}.

In fig.~\ref{fig:tau1nons} we verify that our implementation of
\eq{taufac} does indeed have the correct logarithmic structure in the
$\TaupT \to 0$ limit. We plot the nonsingular difference between the
leading power expansion of the factorisation formula and an exact
fixed-order calculation at NLO$_1$ accuracy (where the subscript
indicates the number of final state partons). We observe that the
double logarithmic plot correctly shows a linear suppression of the
nonsingular remnant, which approaches zero with falling
$\TaupT$. Fig.~\ref{fig:tau1res} shows the resummed $\TaupT$
distribution at NLL$'$ accuracy (for $q_T>10~\mathrm{GeV}$), which
exhibits the characteristic Sudakov peak. In the following, we will
set the value of $(\TaupT)^\cut$ appearing in the 1- and 2-jet \geneva
differential cross sections of
\eqs{1-jet_cross_section}{2-jet_cross_section} to $1 \GeV$, and the
jet radius $R$ appearing in \eq{Tau1pT_distance} to 0.4.

\subsection{Effect of showering}

One of the advantages of using $\TaupT$ instead of $\Tau_1$ as 1-jet
resolution variable is that it facilitates the matching to showers
ordered in transverse momentum. \footnote{In particular the clustering
done according to \eqs{distance}{TaupT} is equivalent to the
anti-$k_T$ algorithm. This greatly reduces the number of shower
histories that get vetoed in the matching procedure.} In this work, we
shower the partonic events with the $p_T$-ordered \pythiaEight
shower~\cite{Sjostrand:2007gs}, after setting the option
\texttt{SpaceShower:dipoleRecoil=on}~\cite{Cabouat:2017rzi,Monni:2020nks}.
This parameter ensures that the recoil momentum of ISR emissions is
distributed within the colour dipole whenever kinematically possible
(for FSR emissions, this is the default behaviour), thus reducing the
impact of the shower on the colour singlet momentum. In previous work,
this was shown to improve the agreement between parton-level and
showered predictions for the color-singlet transverse momentum
\cite{Alioli:2020qrd,Alioli:2021qbf}. The choice of recoil scheme can
significantly affect the accuracy of PSs and may lead to spurious
contributions at NLL \cite{Nagy:2014mqa, Dasgupta:2018nvj,
  Bewick:2019rbu,Dasgupta:2020fwr,Forshaw:2020wrq,Hamilton:2020rcu}. In
this work, however, we take the PS to set the (LL) accuracy of our
predictions and leave a detailed study to future work.

In fig.~\ref{fig:bbHyH}, we show the effect of showering on the sole inclusive and differential quantity for $b\bar{b}\to H$, i.e. the Higgs boson rapidity. We observe that, as expected, the shower does not alter the distribution.

Moving to more exclusive quantities, in figs.\ref{fig:bbHqT} and
\ref{fig:ccHqT} we show the Higgs boson $q_T$ spectrum in $b\bar{b}\to
H$ and $c\bar{c}\to H$ respectively, again comparing parton-level
predictions with those after showering. For completeness, we also show
the $q_T$ spectrum in $b\bar{b}\to H$ with the option
\texttt{SpaceShower:dipoleRecoil=off} (green) in fig.\ref{fig:bbHqT}.
In this case, we observe deviations of up to 10\% in the central value
compared to the partonic results. Nevertheless, the two options
remain consistent within the scale uncertainties. For the more local
recoil scheme, we observe extremely good agreement between our
partonic results shown in blue (which are formally N$^3$LL+NNLO
accurate) and our showered results in orange.

\begin{figure}[t!]
    \centering
    \includegraphics[width=\linewidth]{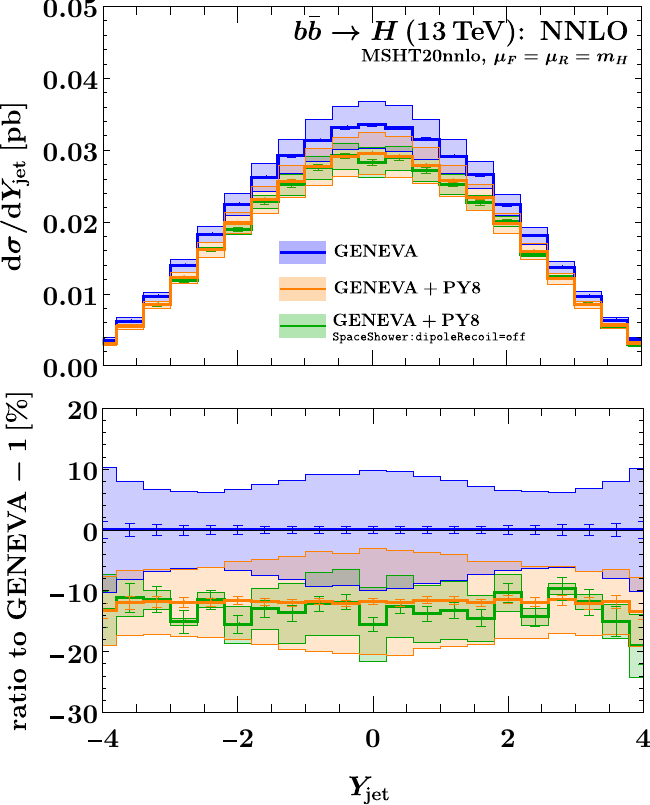}
    \caption{Hardest jet rapidity spectrum in $b\bar{b}\to H$. Results are shown at parton level and after showering.}
    \label{fig:bbHyj}
\end{figure}
\begin{figure}[t!]
    \centering
    \includegraphics[width=\linewidth]{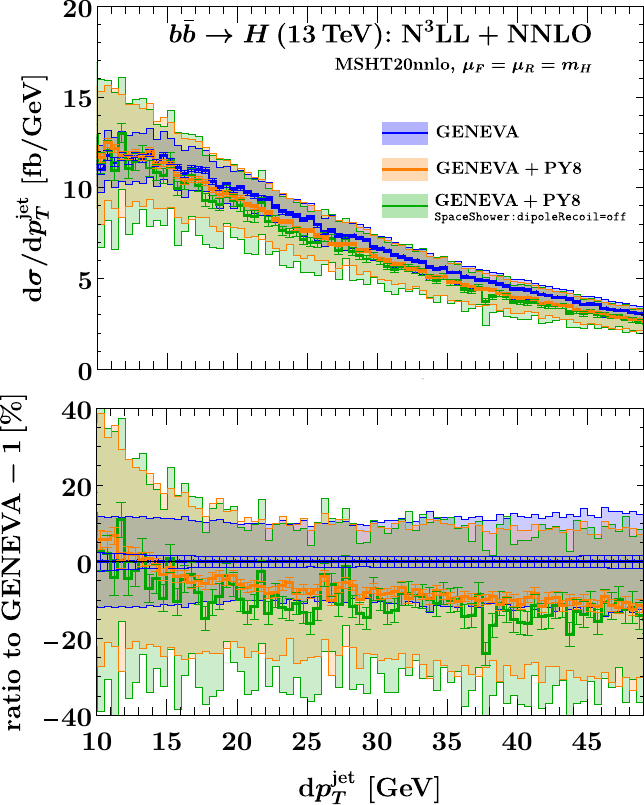}
    \caption{Hardest jet $p_T$ spectrum in $b\bar{b}\to H$. Results are shown at parton level and after showering.}
    \label{fig:bbHpTj}
\end{figure}

In figs.\ref{fig:bbHyj} and \ref{fig:bbHpTj}, we show the rapidity and
transverse momentum of the hardest jet in $b\bar{b}\to H$
respectively. In both the figures, the hardest jet is defined as the
jet with the largest transverse momentum obtained clustering the
final-state partons with an anti-$k_T$ jet algorithm with jet radius
$R = 0.4$. 
Furthermore, in fig.\ref{fig:bbHyj}, jets are required to
have a minimum transverse momentum of $30 \GeV$. The effect of
showering on these exclusive distributions is greater, reaching up to
$\sim 15\%$ on the central value but can be understood from its radiation pattern.
 In the $p_T^{\mathrm{jet}}$ spectrum, we observe a depletion in the high-$p_T^{\mathrm{jet}}$ tail, which arises from additional radiation generated by the shower that redistributes energy among the final-state partons and reduces the transverse momentum of the leading jet. 
This behaviour also impacts the rapidity distribution when imposing the
requirement $p_T^{\mathrm{jet}} > 30~\mathrm{GeV}$ on the hardest
jet. In this region, the showered prediction lies below the partonic
result, which means that fewer events pass the cut after showering,
leading to a reduced overall differential cross section in the rapidity spectrum.
\section{Conclusions}
\label{sec:conc}

In this work we have investigated the use of transverse resolution
variables to achieve NNLO+PS matching within the \geneva framework. In
particular, we combine $q_T$ resummation at N$^3$LL, obtained via an
interface to the \textsc{SCETlib} library, with $\TaupT$ resummation
at NLL$'$. This combination of variables facilitates matching to
transverse momentum ordered showers while also providing high accuracy
predictions for the colour singlet transverse momentum. We have tested
our framework on Higgs boson production in heavy-quark annihilation,
in particular examining the $b\bar{b}\to H$ channel, finding good
agreement with independent fixed order calculations and verifying that
the $q_T$ distribution is numerically preserved after showering, even
though the formal logarithmic accuracy of the showered events remains
limited by the shower LL accuracy.

A possible direction for future work would be to compare the results
of this calculation with those obtained using a different resummation
variable, which could provide useful insights into the estimation of
the theoretical uncertainties. Even though the \geneva framework
permits the use of different resummation variables, the switch between
them has not yet been automated and requires a dedicated
implementation for each process. To get a sense of the expected size
of the effect, one could look at the comparison between $q_T$ and
$\Tau_0$ resummation performed for the Drell-Yan process in
\refcite{Alioli:2021qbf} (see figs.~8-10). In this case, one observes
an $\ord{10\%}$ effect across most of the spectra with larger
differences at small values of the resummation variables (in
particular in the $\Tau_0$ distribution, where they are not covered by
the theoretical uncertainties estimated using scale variations).

Moreover, the methods developed here could be extended to the production of other colour singlet processes. This could for example be useful in the context of $M_W$ extractions -- for example, in \refcite{CMS:2024lrd}, differential distributions for $W$ and $Z$ boson production obtained using the \textsc{MiNNLOps} generator~\cite{Monni:2019whf} are reweighted to N$^3$LL+NNLO accuracy with \textsc{SCETlib}. Extending the generators we have developed here to the Drell-Yan case would remove the need for this reweighting step, since we interface \textsc{SCETlib} directly and reach N$^3$LL+NNLO accuracy in the $q_T$ distribution by construction. 

Another interesting direction to pursue is the extension of the resummation accuracy in $\TaupT$ to NNLL$'$. This would allow matching fixed-order calculations for colour singlet production in association with a hard jet to parton shower algorithms. Although at present the most promising route to this goal seems to be using the normal, invariant mass definition of the one-jettiness (and indeed, both the required fixed-order and resummed calculations for the $Z+j$ process are now implemented in \geneva ~\cite{Alioli:2023rxx,Alioli:2025hpa}), as in the colour singlet case the transverse nature of $\TaupT$ is likely to facilitate matching to transverse momentum-ordered showers. Extension of the resummation accuracy would require the computation of an appropriate 2-loop soft function, which could for example be achieved using \textsc{SoftSERVE}~\cite{Bell:2018oqa,Bell:2020yzz,Bell:2023yso}.

Finally, an important topic in perturbative calculations is the consistent estimation of theoretical uncertainties related to missing higher orders. Approaches based on theory nuisance parameters (TNPs) have shown promise in both resummed and fixed order calculations~\cite{Tackmann:2024kci, Lim:2024nsk} and how similar techniques could be applied to the event generation case is an interesting open question. The fact that a direct application of the TNP approach to the resummed $q_T$ spectrum has already been presented in \refcite{Tackmann:2024kci} perhaps suggests that generators of the kind we have presented in this work (using SCET-based resummation and $q_T$ as resolution variable) may provide a starting point. We leave this to future work.

\acknowledgments
We are grateful to Ciaran Williams for providing the NNLO predictions for $b\bar{b}\to H$ from MCFM which were used in our validation. We thank Simone Alioli, Andrea Banfi, Alessandro Broggio, Johannes Michel and Frank Tackmann for useful discussions, and Thomas Cridge for early collaboration on this project. We also thank  our other \geneva and \scetlib collaborators for their work on the code. This project has received funding from the European Research Council (ERC)
under the European Union's Horizon 2020 research and innovation programme
(Grant agreement 101002090 COLORFREE). MAL was supported by the UKRI guarantee scheme for the Marie Sk\l{}odowska-Curie postdoctoral fellowship, grant ref. EP/X021416/1.

\appendix
\section{Profile scales and scale variation}
\label{app:profilescales}
The canonical boundary scales in $b_T$ space are given by
\begin{align}\label{eq:canonical_scales}
\text{virtuality: }& \quad \mu_H = m_H, \quad \mu_B = b_0/b_T, \quad \mu_S = b_0/b_T,\nn \\
&\hspace{2.1cm} \quad \mu_f = b_0/b_T,
\quad \mu_0 =  b_0/b_T, \nn \\
\text{rapidity: }& \hspace{2.45cm} \nu_B = m_H,\,  \hspace{0.66cm} \nu_S = b_0/b_T
\,,\end{align}
where $b_0\equiv 2 e^{-\gamma_E}\approx 1.12291$, $\mu_H$, $(\mu_B, \nu_B)$, and $(\mu_S, \nu_S)$ are the boundary scales for the
hard, beam, and soft functions, and $\mu_f$ is the scale at which the PDFs inside the beam functions are evaluated. The rapidity anomalous dimension must also be resummed and $\mu_0$ is its associated boundary scale. Evolution of each of the functions in \eq{tmd_factorization1} from these
scales to common scales $\mu,\,\nu$ resums all canonical $b_T$-space logarithms $\ln^n[(b_0/b_T)/m_H]$. The corresponding resummed result in $q_T$ space is then obtained via inverse Fourier transform.

Choosing scales which are functions of the variable being resummed facilitates deactivation of the resummation in the fixed-order region of phase space. We choose hybrid profile scales, depending on both $b_T$ and $q_T$, as
\begin{align} \label{eq:central_scales}
\mu_H &= \nu_B = \mu_\FO = m_H
\,,\nn\\
\mu_X &= m_H\, \frun\Bigl[ \frac{q_T}{m_H}, \frac{1}{m_H} \mu_*\Bigl(\frac{b_0}{b_T}, \mu_X^\mathrm{min}\Bigr)\Bigr] 
\,, \nn \\
\mu_0 &= \mu_*\Bigl(\frac{b_0}{b_T}, \mu_0^\mathrm{min}\Bigr)
\,,\end{align}
where $\mu_X \in \{\mu_B, \mu_S, \nu_S, \mu_f\}$.
The function $\mu_*$ specifies our nonperturbative prescription, which prevents the beam and soft scales reaching values $1/b_T \lesssim \lqcd$ and allows the inverse Fourier transform to $q_T$ space to be performed. It is given by
\begin{equation}
\mu_*(x,y) = \sqrt{x^2 + y^2}
\end{equation}
$\frun$ is the hybrid profile function given by~\cite{Lustermans:2019plv}
\begin{align}
\frun(x,y)&= 1 + \grun(x)(y-1)
\,,\end{align}
where $\grun(x)$ determines the transition as a function of $x = q_T/m_H$,
\begin{align}\label{eq:def_g_run}
\grun(x) &= \begin{cases}
1 & 0 < x \leq x_1 \,, \\
1 - \frac{(x-x_1)^2}{(x_2-x_1)(x_3-x_1)} & x_1 < x \leq x_2
\,, \\
\frac{(x-x_3)^2}{(x_3-x_1)(x_3-x_2)} & x_2 < x \leq x_3
\,, \\
0 & x_3 \leq x
\,,\end{cases}
\end{align}
with transition points $x_i$ for $i\in\{1,2,3\}$. The parameters $x_1$ and $x_3$ determine the start and end of the transition and $x_2 =(x_1+ x_3)/2$ corresponds to the turning point. We use $[x_1, x_2, x_3] = [0.1, 0.45, 0.8]$ as our central values.

The profile scales are varied as follows:
\begin{align} \label{eq:profile_vars}
\mu_H &= \mu_\FO = 2^{w_\FO}\, m_H
\,, \nn \\
\nu_B &= \mu_\FO \, f_\vary^{v_{\nu_B}}\Bigl(\frac{q_T}{m_H}\Bigr)
\,, \nn \\
\mu_X &= \mu_\FO \, f_\vary^{v_{\mu_X}}\Bigl(\frac{q_T}{m_H}\Bigr) f_\run\biggl[
\frac{q_T}{m_H},
\frac{1}{m_H}\mu_*\Bigl(\frac{b_0}{b_T}, \frac{\mu_X^\mathrm{min}}{2^{w_\FO}f_\vary^{v_{\mu_X}} } \Bigr)
\biggr]\,, \nn \\
&\quad\text{for}\quad \mu_X \in \{\mu_B, \mu_S, \nu_S\}
\,, \nn \\
\mu_f &= 2^{w_F}\, m_H\,  f_\run\biggl[
\frac{q_T}{m_H},
\frac{1}{m_H}\mu_*\Bigl(\frac{b_0}{b_T}, \frac{\mufmin}{2^{w_F}} \Bigr)
\biggr]
\,,\nn\\
\mu_0
&= \mu_*\Bigl( \frac{b_0}{b_T}, \muZeromin \Bigr)
\,.\end{align}

Resummation uncertainties are gauged by varying the exponents $v_{\mu_B}$, $v_{\nu_B}$, $v_{\mu_S}$, and $v_{\nu_S}$ about their central values $v_i=0$ by $\pm1$.
The function 
\begin{align}
f_\vary(x) &= \begin{cases}
2(1-x^2/x_3^2) &  0\leq x \leq x_3/2
\,, \\
1-2(1-x/x_3)^2  & x_3/2 < x \leq x_3
\,, \\
1 & x_3 \leq x
\,,\end{cases}
\end{align}
with $x \equiv q_T/m_H$
controls the size of the variations, ranging from a factor of 2 for $x = 0$ to 1 for $x \geq x_3$. Note that both the resummation itself and the associated resummation uncertainty is deactivated for $q_T \geq x_3 m_H$. The combined resummation uncertainty $\Delta_\res$ is calculated by enveloping 36 variations of suitable combinations of the $v_i$. For details, we refer the reader to \refcite{Ebert:2020dfc}.

For the fixed-order uncertainty $\Delta_\FO$, we vary $\mu_\FO$ by a factor of 2 by taking $w_\FO =\{-1,0,+1\}$ everywhere. A separate uncertainty $\Delta_{\mu_f}$ is related to the DGLAP running of the PDFs, for which we vary the PDF scale $\mu_f$ by taking $w_F =\{-1,0,+1\}$. In the nonsingular and fixed-order cross sections, this corresponds to taking $\mu_f \equiv \mu_F = 2^{w_F} m_H$. The resulting $\Delta_\FO$ and $\Delta_{\mu_f}$ are then given by the maximum envelope of the respective variations. Finally, a matching uncertainty $\Delta_{\mathrm{match}}$ is obtained by varying $x_2$ in the range [0.2,0.6].

The total perturbative uncertainty for exclusive quantities is obtained by quadrature sum of the individual contributions, viz.
\begin{align}
\Delta_{\mathrm{total}}^2 =\Delta_\FO ^2 + \Delta_{\mathrm{res}}^2 + \Delta_{\mu_f}^2 + \Delta^2_{\mathrm{match}}.
\end{align}
For inclusive quantities, we instead take the conventional 7-point scale variation band.

\section{Plus distributions}
\label{app:plusdistrib}
Following \refcite{Ligeti:2008ac}, we define plus distributions of dimensionless arguments as 
\begin{align}
\mathcal{L}_n(x)&\equiv \left[\frac{\Theta(x)\ln^nx}{x}\right]_+\nn\\
&=\lim_{\beta\to 0}\left[\frac{\Theta(x-\beta)\ln^n x}{x}+\delta(x-\beta)\frac{\ln^{n+1}\beta}{n+1}\right]\,,
\end{align}
\begin{align}
\mathcal{L}^a(x)&\equiv \left[\frac{\Theta(x)}{x^{1-a}}\right]_+\nn\\
&=\lim_{\beta\to 0}\left[\frac{\Theta(x-\beta)}{x^{1-a}}+\delta(x-\beta)\frac{x^a-1}{a}\right]\,.
\end{align}
Solutions to the evolution equations for $\Tau_1^{p_T}$ are conveniently expressed in terms of the function $\mathcal{V}_a(x)$, which is defined as
\begin{align}
\mathcal{V}_a(x)&\equiv \frac{e^{-\gamma_E a}}{\Gamma(1+a)}\left[a\mathcal{L}^a(x)+\delta(x)\right]\,,\nn\\
\mathcal{V}_a(k,\mu)&\equiv \frac{1}{\mu}\mathcal{V}_a\left(\frac{k}{\mu}\right)\,.
\end{align}
The $\mathcal{V}_a$ satisfy a number of useful identities, 
\begin{align}
\int \mathrm{d}k' \mathcal{V}_a(k',\mu)\mathcal{V}_b(k-k',\mu)&=\mathcal{V}_{a+b}(k,\mu)\,,\\
\mathcal{V}_a(k,\mu)&=\left(\frac{\mu'}{\mu}\right)^a\mathcal{V}_a(k,\mu')\,,\\
\mu \frac{\mathrm{d}}{\mathrm{d}\mu}\mathcal{V}_a(k,\mu)&=-a\mathcal{V}_a(k,\mu)\,.
\end{align}

\begin{figure}[ht!]
    \centering
    \includegraphics[width=\linewidth]{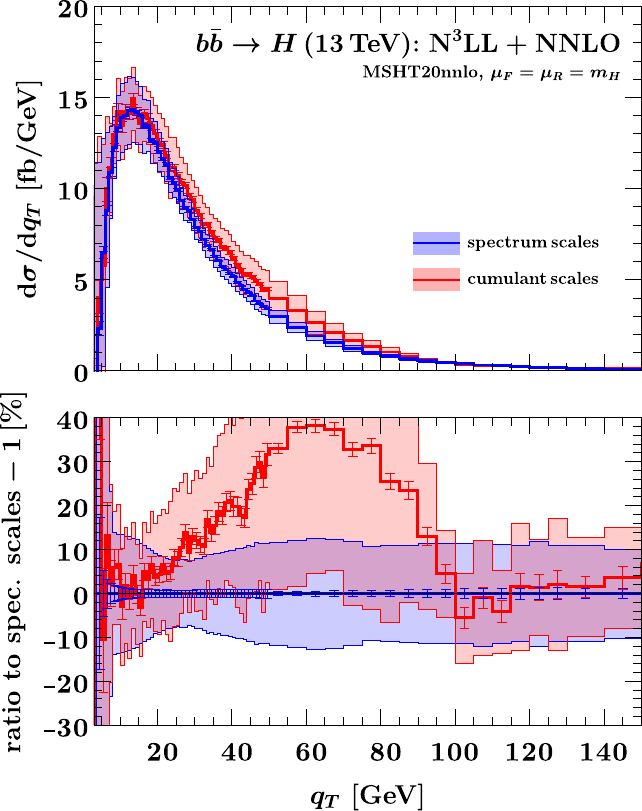}
    \caption{Parton-level Higgs boson $q_T$ spectrum in $b\bar{b}\to H$ for spectrum and cumulant scale settings.}
    \label{fig:bbHqT_q}
\end{figure}

\section{One-loop expressions for $\Tau_1^{p_T}$ soft, beam and jet functions}
\label{app:functions}
Each of the soft, beam and jet functions admits a perturbative expansion in $\as$ of the form
\begin{align}
F=\sum_{n=0}^\infty \left(\frac{\as}{4\pi}\right)^nF^{(n)}
\end{align}
for $F\in\{S,B,J\}$. Furthermore, the beam function can be written as a convolution between the normal parton distribution functions and perturbative kernels as
\begin{align}
B_i\left(k,x,\mu,\frac{\nu}{\omega}\right)=\sum_m\int^1_x\frac{\mathrm{d}z}{z}\mathcal{I}_{im}\left(k,z,\mu,\frac{\nu}{\omega}\right)\,f_m\left(\frac{x}{z},\mu\right)
\end{align}
where it is the kernels $\mathcal{I}$ that possess an expansion in $\as$. 

The one-loop expressions necessary for the $\Tau_1^{p_T}$ beam function at NLL$'$ are given by
\begin{align}
\mathcal{I}_{im}^{(1)}(z) &=
\delta_{im} \delta(1-z) \, L^\mu_B \left(2 C_i\Gamma_0 L^\nu_B+\gamma^{i(0)}_{\mu,B}\right) \nn \\
&\qquad-2 L^\mu_B P_{im}^{(0)}(z)
+I_{im}^{(1)}(z)
\,, \\
  I_{gq}^{(1)}(z)&=z\, ,\\
  I_{qq}^{(1)}(z)&=1-z\, ,\\
  I_{qg}^{(1)}(z)&=2z(1-z)\, ,
\end{align}
where the logarithms are defined as 
\begin{equation}
L_B^\mu = \ln \frac{\mu}{k}
\,, \quad
L_B^\nu = \ln \frac{\nu}{\omega}
\,.\end{equation}
The quark and gluon jet functions at one loop are given by
\begin{align}
    J^{(1)}_q(k,\mu,\nu) &= C_F\Bigg[\left(7-\frac{2\pi^2}{3}-6\log 2\right)\delta(k)\nn \\
    & \qquad +\left(6+8\log\frac{\nu}{Q}\right)\mathcal{L}_0\left(k,\frac{\mu}{Q}\right)\Bigg]
\end{align}
and 
\begin{align}
    J^{(1)}_g(k,\mu,\nu) &= \Bigg[C_A\left(\frac{25}{12}-\frac{2\pi^2}{3}\right)+\beta_0\left(\frac{17}{12}+2\log 2\right)\Bigg]\delta(k)\nn \\
    & \qquad +\left[2\beta_0+8C_A\log\frac{\nu}{Q}\right]\mathcal{L}_0\left(k,\frac{\mu}{Q}\right)\,.
\end{align}
The soft function at one-loop order was calculated in \refcite{Bertolini:2017efs}. While the full expression is rather lengthy (and can be found in that reference after setting $\beta=\gamma=1$), the term proportional to $\delta(k_S)$ is given by
\begin{align}
s^{(1)}_{\kappa} &= -\left(C_a+C_b+C_J\right)\frac{\pi^2}{6}\nn \\
&\qquad -2C_J\log^2 R_J\left[2+\Theta(R_J-1)\right]
\end{align}
where we have neglected power-suppressed terms in $R_J^2$. These have been shown to have at worst a moderate numerical impact, even for values of the jet radius $R_J$ up to 1.5~\cite{Bertolini:2017efs}.

\section{Splitting mappings and splitting functions}

We report in this section the explicit expressions for the splitting
mappings $\bar{\Phi}_N(\Phi_{N+1})$ and splitting functions
$\mathcal{P}_{N \to N+1}(\Phi_{N+1})$ used to obtain the results
presented in this paper. If we call $i$ and $j$ the two closest
partons in the phase space point $\Phi_{N+1}$ according to some
distance measures, and $k$ the parton obtained by clustering $i$ and
$j$ together in the phase space point $\bar{\Phi}_N(\Phi_{N+1})$, all
the splitting functions used in the paper take the form
\begin{eqnarray}
  \label{Altarelli-Parisi_splitting_functions}
  && \!\!\!\! \mathcal{P}_{N \to N+1}(\Phi_{N+1}) =
  \\
  && \!\!\!\! \left\{
  \begin{array}{l l}
    \displaystyle f_j^a\!\left(x_a,\mu_F\right)
    f_{p_b}^b\!\left(x_b,\mu_F\right) \mathcal{P}_{kj}^{\rm
      ISRA}\!\left(\Phi_{N+1}\right) & \mbox{if $k$ is \textit{ISRA}}
    \\ [3mm]
    \displaystyle f_{p_a}^a\!\left(x_a,\mu_F\right)
    f_j^b\!\left(x_b,\mu_F\right) \mathcal{P}_{kj}^{\rm
      ISRB}\!\left(\Phi_{N+1}\right) & \mbox{if $k$ is \textit{ISRB}}
    \\ [3mm]
    \displaystyle f_{p_a}^a\!\left(x_a,\mu_F\right)
    f_{p_b}^b\!\left(x_b,\mu_F\right) \mathcal{P}_{jk}^{\rm
      FSR}\!\left(\Phi_{N+1}\right) & \mbox{if $k$ is \textit{FSR}},
  \end{array}
  \right.
  \nonumber
\end{eqnarray}
where $f_p^h(x_h,\mu_F)$ is the PDF for the parton $p$ in the hadron
$h$ ($a$ or $b$) with fraction of momentum $x_h$ at factorization
scale $\mu_F$, and $\mathcal{P}_{kj}^{\rm ISRA}(\Phi_{N+1})$,
$\mathcal{P}_{kj}^{\rm ISRB}(\Phi_{N+1})$ and $\mathcal{P}_{jk}^{\rm
  FSR}(\Phi_{N+1})$ are three functions whose explicit expressions
depend on the splitting mappings and are provided in the following
sections. The labels $p_a$ and $p_b$ identify the partons extracted
from the first and second beam (if those are not the partons that
undergo the splitting, their flavour will not change between
$\Phi_{N+1}$ and $\bar{\Phi}_N(\Phi_{N+1})$). To obtain the results
presented in this paper, we have used the distance measures given in
\eq{Tau1pT_distance}. Since \eq{Tau1pT_distance} does not distinguish
emissions coming from different ISR legs, we call ISRA the emissions
with positive rapidity and ISRB those with negative rapidity in the
reference frame obtained by longitudinally boosting the laboratory
frame to the frame where the color singlet has zero rapidity.

\subsection{$1 \to 0$ mapping}

The $1 \to 0$ mapping $\bar{\Phi}_0(\Phi_1)$ used in this paper is the
one that preserves
\begin{enumerate}
\item The mass of the color singlet.
\item The rapidity of the color singlet.
\end{enumerate}
The resolution variable used for the $0 \to 1$ splitting is the
transverse momentum $q_T$ of the color singlet. Furthermore, after
calling $p_1$ the momentum of the emitted parton $i$, in the ISRA
case, we define
\begin{equation}
  z = \frac{\sqrt{S} \bar{x}_a}{\sqrt{S} \bar{x}_a + p_1^-},
\end{equation}
while in the ISRB case
\begin{equation}
  z = \frac{\sqrt{S} \bar{x}_b}{\sqrt{S} \bar{x}_b + p_1^+},
\end{equation}
where $\bar{x}_a$ and $\bar{x}_b$ are the fractions of momentum
carried by the two initial partons in the $\bar{\Phi}_0(\Phi_1)$ phase
space point, and $\sqrt{S}$ is the collider energy. We notice that, in
the collinear limit, the above definition of $z$ coincides with the
energy ratio between the partons $k$ and $j$. At this point, we can
use a single function
\begin{equation}
  \mathcal{P}_{kj}^{\rm ISR}\!\left(\Phi_1\right) =
  \frac{8\pi\as\!\left(\mu_R\right)}{q_T^2} \left(1-z\right) P^{\rm
    AP}_{jk}\!\left(z\right).
\end{equation}
to cover both the ISRA and ISRB cases, with the implicit understanding
that the definition of $z$ is different between the two cases. The
functions are the azimuthally-integrated unregulated Altarelli--Parisi
splitting functions
\begin{eqnarray}
  P^{\rm AP}_{qq}\!\left(z\right) &=& C_F \, \frac{1+z^2}{1-z}
  \\
  P^{\rm AP}_{qg}\!\left(z\right) &=& C_F \,
  \frac{1+\left(1-z\right)^2}{z}
  \\
  P^{\rm AP}_{gq}\!\left(z\right) &=& T_F \left[z^2 +
    \left(1-z\right)^2\right]
  \\
  P^{\rm AP}_{gg}\!\left(z\right) &=& 2C_A \left[\frac{z}{1-z} +
    \frac{1-z}{z} + z\left(1-z\right)\right].
\end{eqnarray}

\subsection{$2 \to 1$ ISR mapping}

We define the $2 \to 1$ ISR mapping $\bar{\Phi}_1(\Phi_2)$ as the one
that preserves
\begin{enumerate}
\item The momentum $q$ of the color singlet.
\item The rapidity $\eta_1$ of the final-state parton that does not
  come from the splitting.
\end{enumerate}
The resolution variable used for the $1 \to 2$ ISR splitting is
\begin{equation}
  \Tau_1 = p_{2T},
\end{equation}
where $p_2$ the momentum of the emitted parton $i$, and, similarly to
what we did above, in the ISRA case, we define
\begin{equation}
  z = \frac{\sqrt{S} \bar{x}_a}{\sqrt{S} \bar{x}_a + p_2^-}
\end{equation}
while in the ISRB case
\begin{equation}
  z = \frac{\sqrt{S} \bar{x}_b}{\sqrt{S} \bar{x}_b + p_2^+},
\end{equation}
where $\bar{x}_a$ and $\bar{x}_b$ are the fractions of momentum
carried by the two initial partons in the $\bar{\Phi}_1(\Phi_2)$ phase
space point, and $\sqrt{S}$ is the collider energy. Similarly to what
we did above, we then define
\begin{equation}
  \mathcal{P}_{kj}^{\rm ISR}\!\left(\Phi_2\right) =
  \frac{8\pi\as\!\left(\mu_R\right)}{\Tau_1^2} \left(1-z\right) P^{\rm
    AP}_{jk}\!\left(z\right).
\end{equation}

\subsection{$2 \to 1$ FSR mapping}

We define the $2 \to 1$ FSR mapping $\bar{\Phi}_1(\Phi_2)$ as the one
that preserves
\begin{enumerate}
\item The momentum $q$ of the color singlet.
\item The rapidity $\eta_1$ of the final-state parton with the largest
  transverse momentum.
\end{enumerate}
The resolution variable used for the $1 \to 2$ FSR splitting is
\begin{equation}
  \label{eq:FSR_Tau1}
  \Tau_1 = \sqrt{\frac{p_{T2}}{p_{T1}} \left(p_1+p_2\right)^2},
\end{equation}
where $p_2$ is the momentum of the emitted parton $i$ and $p_1$ that
of the other final-state parton $j$. Furthermore, we define
\begin{equation}
  z = \frac{p_{1T}}{p_{1T}+p_{2T}}.
\end{equation}
We notice that, in the collinear limit, the above definition of $z$
coincides with the energy ratio between the partons $j$ and $k$. This
is the first FSR splitting we encounter in the calculation. We define
\begin{equation}
  \mathcal{P}_{kj}^{\rm FSR}\!\left(\Phi_2\right) =
  \frac{8\pi\as\!\left(\mu_R\right)}{\Tau_1^2} \frac{1-z}{z} P^{\rm
    AP}_{kj}\!\left(z\right).
\end{equation}

\subsection{Second $2 \to 1$ FSR mapping}

In \geneva, we make use of two different kind of splitting
mappings. The $\bar{\Phi}_N(\Phi_{N+1})$ mappings are used above the
$r_N$ resolution cut to spread the $r_N$ resummed distribution over
the differential $\Phi_{N+1}$ events, while the
$\tilde{\Phi}_N(\Phi_{N+1})$ mappings are used below the $r_N$
resolution cut to subtract the infrared divergences of the
$\Phi_{N+1}$ events at small $r_N$. Since the two mappings are used on
complementary subsets of the $d\Phi_{N+1}$ phase space, they do not
need to coincide, although they did in fact coincide in all the
previous \geneva implementations. In this paper, however, using the
same FSR mappings above and below the $\Tau_1$ resolution cut would
have resulted in significant complications in their computational
implementations. For this reason, we made use of a different FSR $2
\to 1$ $\tilde{\Phi}_1(\Phi_2)$ mapping, defined as the one that
preserves
\begin{enumerate}
\item The momentum $q$ of the color singlet.
\item The rapidity $\eta_{12}$ of the final-state jet (given by the
  two clustered partons in $\Phi_2$).
\end{enumerate}
This choice allows for a simpler treatment of the standard $\xi$, $y$
and $\phi$ variables used when dealing with the subtraction of the
infrared diveregences, since they are typicallly defined in terms of
energies and angles of the emitted parton $p_2$ with respect to the
emitter $p_1+p_2$. On the other hand, the $\bar{\Phi}_1(\Phi_2)$
mapping proves more convenient when the splitting function is written
in terms of $\Tau_1$, as defined in \eq{FSR_Tau1}, which depends
separately on $p_{1T}$ and $p_{2T}$.

\subsection{Impact of the choice of the splitting mapping}

In the \geneva formalism, the choice of the splitting mapping and the
resolution variables affects the way the so-called splitting functions
are implemented (see section 3.1 of \refcite{Alioli:2023har} for the
technical details). These are used to determine the event weights in a
way that guarantees that the generated events are distributed
according to the resummed spectrum in the given resolution variable
(in this case, the transverse momentum of the Higgs boson). The way
\geneva achieves this requires dividing each event weight by the cross
section (approximated in the collinear limit) at a given underlying
phase space point $\Phi_N$ (determined by the splitting mapping) and
value of the resolution variable $r_N$. This approach requires
computing the integration limits for the splitting variables at fixed
$\Phi_N$ and $r_N$, and the feasibility of such a calculation
significantly limits the number of new mappings that can be
implemented.

\begin{figure}[t!]
	\centering
	\includegraphics[width=0.48\textwidth]{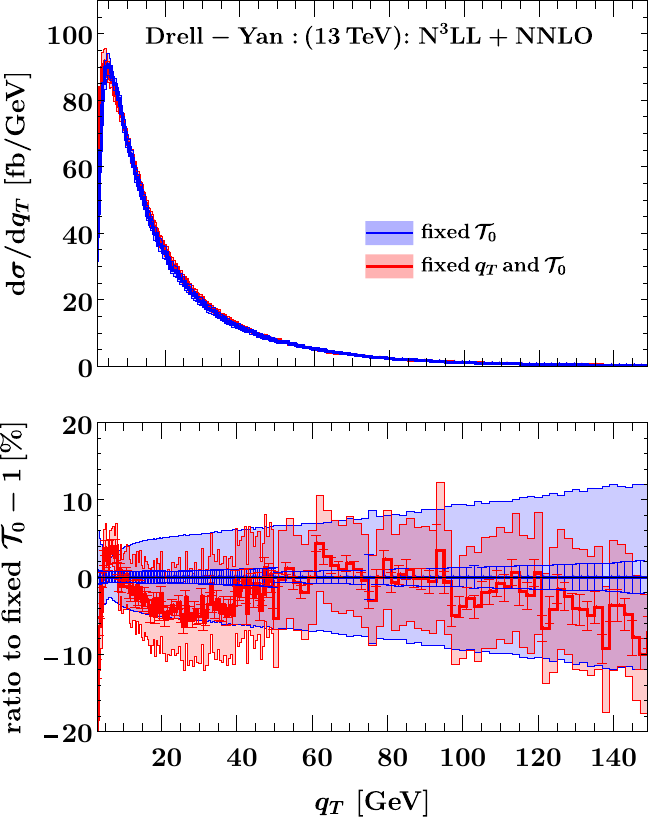}
\caption{Transverse momentum spectrum of the vector boson for the
  Drell-Yan process with different splitting mappings.}
\label{fig:DY}
\end{figure}

In the work presented in this paper, the expression of the
$\mathcal{T}_1^{p_T}$ resolution variable makes the implementation of
alternative splitting mappings very complicated. To nevertheless
investigate the expected uncertainty associated with the choice of the
mapping, we have performed a dedicated study using the existing
implementation of the Drell-Yan process, where the standard
$N$-jettiness (see eq.~(12) of the draft) was used as resolution
variable and two different splitting mappings are currently
available. The results of the comparison are shown in
fig.~\ref{fig:DY}. Both the splitting mappings here preserve the
resolution variable $\Tau_0$, but one of them (corresponding to the
red distributions in the figure) also preserves the transverse
momentum of the vector boson. From the ratio plot in the lower pane,
we can see that the numerical impact associated with the choice of the
different mappings appears to be below 5\% across the entire spectrum
and within the theoretical uncertainty estimated through scale
variations.

\section{Effect of scale setting choices on the $q_T$ spectrum}
\label{app:cumulant_scales}
In fig.\ref{fig:bbHqT_q} we show the Higgs boson $q_T$ spectrum in $b\bar{b}\to H$ for spectrum (blue) and cumulant (red) scale settings. These different scale choices were discussed in detail in \sec{FOval}. We observe that both spectra are in good agreement in the peak and the tail region. In particular for the tail of distribution, this behaviour is expected as both distributions are matched to the same fixed-order prediction and therefore must agree in the fixed-order limit. In the transition region, however, the higher order terms have a significant effect which can be as large as 40\%. The large size of this effect is peculiar to the $b\bar{b}\to H$ process, and its origin is discussed in detail in \refcite{Cal:2023mib}.

\section{Flavour structure of \textsc{Geneva}}
The \geneva method is sufficiently general that a number of different choices for the resolution variables $r_0$ and $r_1$ may be made, each resulting in an NNLO+PS accurate generator. To date, choices for $r_0$ include the transverse momentum of the colour singlet $q_T$, the $0-$jettiness $\Tau_0$ and the hardest jet transverse momentum $p_{1,T}^j$, while choices for $r_1$ include the conventional $1-$jettiness $\Tau_1$, the second jet transverse momentum $p_{2,T}^j$ and, as of this work, the $1-$jettiness with transverse momentum measures $\Tau_1^{p_T}$. It is clear that the number of different permutations now available may cause confusion when one wishes to refer to a specific implementation of \geneva. Inspired by the origins of the term `flavour' in particle physics (due to Gell-Mann), we therefore propose a nomenclature which should serve to clarify exactly which choice of resolution variables has been made in a given case. This appears in tab.~\ref{tab:flavours}. 

\begin{table}[]
\centering
\begin{tabular}{c||cc | c}
Flavour\,       & $r_0$ & $r_1$  & \, Reference\\ \hline \hline
Vanilla (V)\, & \,$\Tau_0$     & $\Tau_1$ & \cite{Alioli:2015toa}  \\ \hline
Raspberry ripple (RR)\, &\, $q_T$     & $\Tau_1$ & \cite{Alioli:2021qbf}    \\ \hline
Tutti frutti (TF)\,& \,$p_{1,T}^j$     & $p_{2,T}^j$ & \cite{Gavardi:2023aco}  \\ \hline
Mint choc chip (MCC)\,& \,$q_T$     & $\Tau_1^{p_T}$  & This work
\end{tabular}
\caption{Flavours of \geneva implementation.}
\label{tab:flavours}
\end{table}

\bibliographystyle{apsrev4-1}
\bibliography{bbh}
\end{document}